\documentclass[12pt]{article}
\usepackage{nomencl}
\usepackage{setspace}
\usepackage{graphicx}
\usepackage{adjustbox}
\usepackage[utf8]{inputenc}
\usepackage[T1]{fontenc}
\usepackage{amsmath}
\usepackage{amssymb}
\usepackage{array}
\usepackage{bibentry}
\usepackage{caption}
\usepackage{commath}
\usepackage{enumitem}
\usepackage{esint}
\usepackage{pdflscape}
\usepackage{siunitx}
\usepackage{graphicx}
\usepackage[colorlinks=true, linkcolor=blue, citecolor=black, urlcolor=black]{hyperref}

\usepackage{ifthen}
\usepackage{lipsum}
\usepackage{setspace}
\usepackage{subcaption}
\usepackage{tcolorbox}
\tcbuselibrary{breakable}
\usepackage{tikz}
\usepackage{xcolor}
\usepackage{siunitx}
\usepackage{array}
\usepackage{pifont}
\usepackage[authoryear,round]{natbib} 

\usepackage{fullpage} 

\usepackage{amsmath}
\newcommand{\DEFN}{\equiv}

\newcommand{\D}{\mathrm{d}}

\renewcommand{\D}{\mathrm{d}}

\usepackage{setspace}

\usepackage{tabularx}                          
\usepackage{natbib}
\usepackage{apalike}
\usepackage[font={small}]{caption}
\usepackage{amsmath}                            
\usepackage{amssymb} 
\usepackage{graphicx}                           
\usepackage{gensymb}
\usepackage{rotating}

\usepackage{url,lineno,microtype,subcaption}
\usepackage{authblk}

\usepackage{fancyhdr}

\usepackage{booktabs}
\usepackage{rotating}

\usepackage{pifont}

\makenomenclature

\begin{document}

\title {Experimental study on gravity currents flowing on heated walls}

\author[1,*]{Stefano Lanzini}
\author[1]{Massimo Marro}
\author[1]{Mathieu Creyssels}
\author[1]{Alexandre Azouzi}
\author[1,2]{Pietro Salizzoni}

\font\myfont=cmr10 at 11pt

\affil[1]{\myfont  Ecole Centrale de Lyon, CNRS, Universite Claude Bernard Lyon 1, INSA Lyon,
LMFA, UMR5509, 69130 Ecully, France}

\affil[2]{\myfont Department of Environmental, Land and Infrastructure Engineering (DIATI), Politecnico di Torino, Corso Duca degli Abruzzi~24, 10129 Turin (TO), Italy.}
\affil[*]{\myfont Corresponding author: stefano.lanzini@ec-lyon.fr}

\date{Preprint submitted to Experiments in Fluids, Oct. 2025}

\maketitle


\pagebreak

\begin{abstract}
    We present an experimental study on steady gravity currents advancing along a heated wall. The current is generated by a mixture of air and carbon dioxide continuously supplied at the channel inlet. To have a complete point-wise characterization of the flow, simultaneous high-frequency measurements of two velocity components, \textnormal{CO}$_2$ concentration, and temperature are performed. An experimental protocol is presented to reconstruct the local fluid density and to estimate turbulent vertical and horizontal fluxes of \textnormal{CO}$_2$, temperature, and buoyancy. 
The reliability of both the flow measurements and of the estimate of convective heat flux exchanged at the wall is assessed through integral balances of \textnormal{CO}$_2$ mass, enthalpy, and buoyancy, performed at different distances from the source. Three wall-heating conditions are considered: an adiabatic case, a moderately heated case, and a strongly heated case. 

In the heated experiments, a convectively unstable boundary layer forms near the wall, capped by a stably stratified region. The influence of this condition on the first- and second-order flow statistics profiles is examined.
Although wall heating influences the vertical shear, the Brunt–Väisälä frequency, and both shear and buoyancy production of turbulent kinetic energy within the stably-stratified region characterized by an almost constant vertical gradient of streamwise velocity, neither the gradient Richardson number nor the flux Richardson number exhibits a clear trend in this region with the imposed wall heat flux.
\end{abstract}{}



\section{Introduction}
The interaction between airflows in the lower atmosphere and the Earth's surface gives rise to different flow topologies.  
Some of these phenomena, such as sea breezes \citep{Simpson1977}, katabatic winds \citep{Fernando2010}, and thunderstorm outflows \citep{Mueller1987}, can be accurately modeled as gravity currents of relatively cold dense air \citep{Simpson1982}. Laboratory models of such geophysical flows still play an important role in their understanding.  They allow the key flow variables to be controlled and provide access to the measurements statistics from many experiments performed under repeatable conditions, which are hardly attainable in field measurements. In addition to a direct inquiry into the physics of the flow, the dataset produced by laboratory experiments can also be used as a reference to validate and improve numerical models.
The majority of experimental studies concerning flows developing in the lower atmosphere focus on the case of neutral ambient stratification with adiabatic wall conditions; this is the case of studies of atmospheric boundary layers (including the phenomena associated with scalar dispersion within them) as well as of gravity currents. 
In laboratory experiments, variations of conditions at the surface have mainly concerned a varying roughness shape and topology both for atmospheric boundary layers (e.g. \cite{Cheng2002,Castro2006,Castro2007,Southgate-Ash2025}) and -- excluding studies on the influence of bed shape and varying slopes -- gravity currents studies (e.g \cite{Negretti2008, Nogueira2013,Cenedese2018,Jiang2018,Maggi2022,maggi2025}).
Significant efforts have been made to simulate stable and unstable stratification conditions of atmospheric boundary layers. These conditions can be achieved employing water tanks, where ambient density gradients are created either by salinity or temperature gradients (see the review of \cite{YUAN2011}), or by means of thermally stratified wind tunnels (\cite{Webster1964, OGAWA1981, Lienhard1989, MarucciCarpentieriHayden2018, Marucci2020}).
Gravity currents have been extensively investigated using water tanks where the increased density driving the current is induced by an excess of salinity. A number of experimental studies have focused on the influence of a stably stratified ambient environment, given the strong relevance of this condition to both oceanographic and atmospheric applications \citep{Baines2001a,MAXWORTHY2002, Baines_2005,  Longo2016,Chiapponi2018,Dai2021}.\\
On the other hand, unstable conditions at the bottom wall, caused, for example, by wall heating, have not been investigated.
This condition is indeed relevant for atmospheric gravity currents. As an example, the sea breeze current can be significantly influenced by the Earth's surface sensible heat fluxes or by interactions with urban heat islands generated by cities \citep{Simpson1977,CenedeseMonti2003}. The development of a convectively unstable boundary layer within the inland-propagating sea breeze significantly enhances the vertical turbulent transport of pollutants emitted at the surface level. The vertical extent of this mixing can be capped by the layer of unmodified marine air, which can act as a stable interface and inhibit further mixing into the external atmosphere \citep{Lu1994,Miller2003}. Pollutants can be trapped within the sea breeze and can be advected horizontally over long distances. Also, katabatic winds can flow over a warm surface when they reach coastal zones and propagate above lakes and oceans.

To make a step forward in the experimental characterization of these currents, we present a new experimental setup to investigate statistically stationary dense gravity currents, generated by a mixture of air and carbon dioxide, that flow on a heated wall. The goal is to present first- and second-order statistics profiles of different quantities of interest and their evolution for increasing distances from the source, considering different heating intensities.  
The inclusion of heat flux at the wall increases the complexity of the experimental setup and of the experimental techniques required to characterize this flow.  To measure the local flow density and the turbulent flux of different scalars, simultaneous measurements of two velocity components, CO$_2$ concentration, and temperature are required.  

Simultaneous high-frequency measurements of velocity coupled with two different scalars have been seldom reported in the literature. To study multiple scalar mixing, \cite{Sirivat1982} and, more recently, \cite{Hewes_2022} performed simultaneous point-wise measurements of helium concentration, temperature, and one velocity component using three appropriately calibrated and compensated thermal anemometry probes. In experiments in a stratified wind tunnel, \cite{MARUCCI2019a} \cite{Marucci2020} and \cite{Marucci2020b} evaluated passive scalar dispersion in stably and unstably stratified boundary layers, doing simultaneous measurements of two-velocity components with Laser-Doppler Velocimetry, concentration with fast flame ionization detector (FID), and temperature with cold-wire (constant current anemometer, CCA). In this work, the wall-normal and wall-tangent velocity components are measured with Laser-Doppler velocimetry (LDV), and CO$_2$ concentration and temperature are measured with FID and CCA, respectively. The velocity measurements are coupled with simultaneous measurements of two scalars that contribute oppositely to the fluid density. An experimental protocol that allows the determination of both vertical and horizontal buoyancy fluxes by exploiting simultaneous LDV–FID–CCA measurements is presented.

It is worth noting that a major advantage of considering a heated air-CO$_2$ mixture is that no double-diffusive instabilities (like salt-fingering) are expected to develop in the flow. 
These phenomena -- buoyancy-driven motions occurring when two scalars give opposing contributions to vertical density gradient and lead to instabilities even in otherwise stably stratified systems -- require that the two scalars diffuse at substantially different rates (as in the case of salt and temperature in the ocean) \citep{Turner1974}. 
Since the air thermal diffusivity ($\alpha_0=2.12\times 10^{-5}$ m$^2$/s at 20$^\circ$) is very close to the molecular diffusivity of CO$_2$ in air ($D_{\text{CO}_2}=1.61\times 10^{-5}$ m$^2$/s at 20$^\circ$), the occurrence of double-diffusive instabilities is not expected in our experiments. 
This feature is particularly relevant for the present study, as the experiment aims to primarily model gravity currents characterized by a single scalar --- the temperature --- for which double-diffusive instabilities cannot develop \citep{Turner1974}.

The manuscript is structured as follows. The description of the experimental facility and the heating wall, and a characterization of the convection intensity are presented in Section \ref{sec:CH1facility}. The measurement techniques, the solution adopted to address the metrological problem related to the presence of seeding particles and CO$_2$, and the protocol to obtain the turbulent fluxes of the different scalars are discussed in Section \ref{sec:CH1measurements}. Section \ref{sec:CH1inlet} presents the experimental conditions, and the results are discussed in Section \ref{sec:CH1results}. The results section presents the first- and second-order statistics profiles and integral balances of CO$_2$ mass, excess of enthalpy with respect to the source, and buoyancy, performed at different distance from the source. These integral balances serve to validate the flow measurements and the estimate of the convective heat flux supplied to the gravity current, and to verify its spatial homogeneity. Conclusions are reported in Section \ref{sec:CH1conclusion}.

\section{The facility}\label{sec:CH1facility}
\begin{figure}
\centering
\includegraphics[width=1\textwidth]{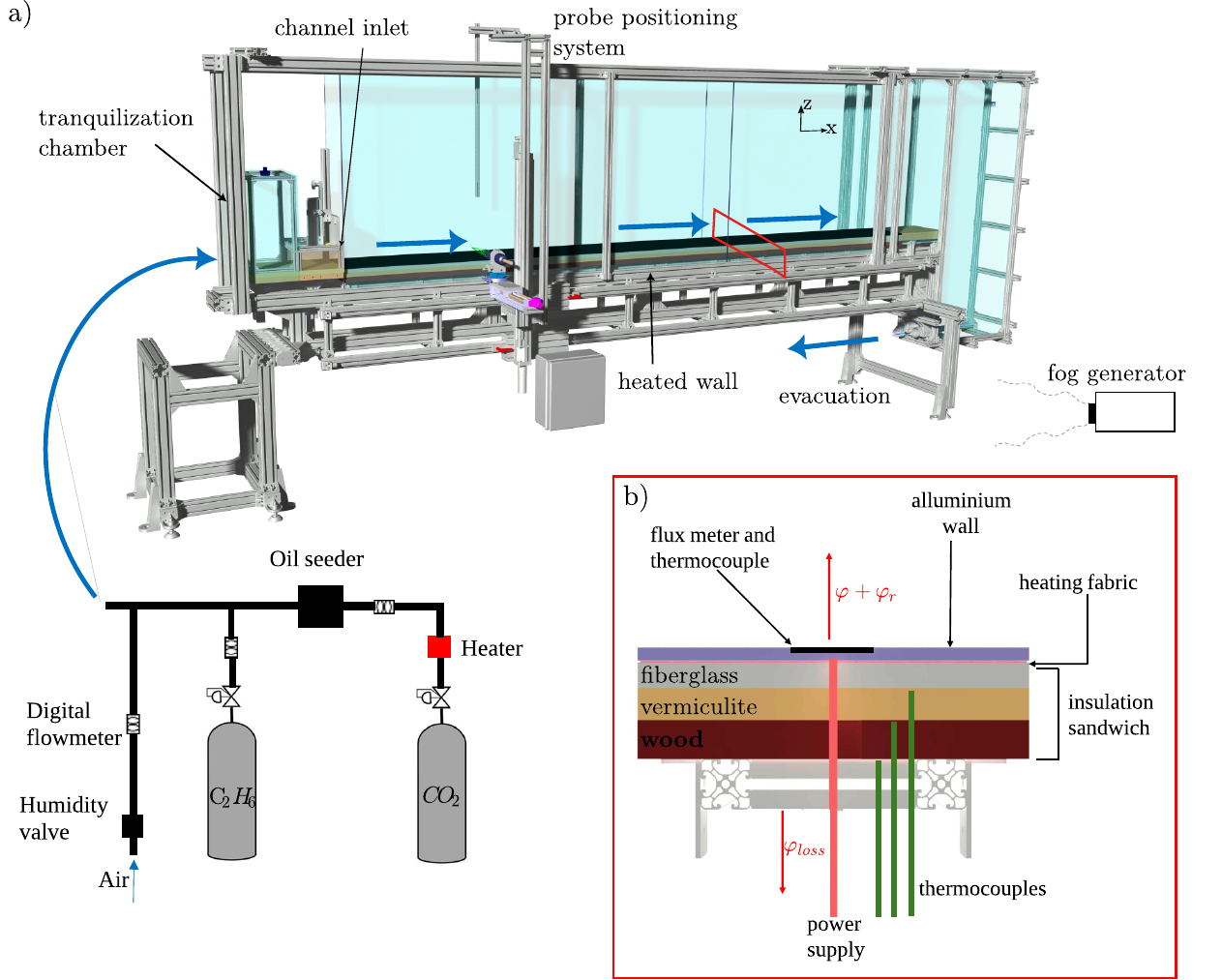}
\caption{(a) Schematic of the experimental set-up. (b) Detail of the heating wall.}
\label{fig:CH1sketch}
\end{figure}
The experimental facility (Figure \ref{fig:CH1sketch}a) was designed and built at the Laboratoire de Mécanique des Fluides et d'Acoustique  (LMFA) at the École Centrale de Lyon (France).
The installation consists of a 4.3-meter-long and 30-cm-wide rectangular channel, featuring a bottom wall that can be heated. The side walls, of height 1.2 m and thickness 6 mm, are made of tempered glass, which can maintain structural integrity at temperatures up to 100 \,°C. The channel is open at the top, allowing ambient air to enter the flow domain.
The dense gravity current is created by a mixture of air and carbon dioxide that is introduced into a tranquilization chamber and then at the channel inlet with a constant mass flux.  
Ethane (C$_2$H$_6$), used as a tracer for CO\textsubscript{2} concentration measurements (Section \ref{subsec:CH1FID}), is also added to the mixture. The volumetric flow rate for each injected gas is regulated by digital mass flow controllers (Alicat Scientific MC-Series), which integrate a proportional–integral–derivative controller and have an accuracy of $\pm 3\%$ \citep{Vidali2022}. CO$_2$ and C$_2$H$_6$ are delivered from pressurized gas bottles, whereas air is supplied from the laboratory’s compressed air line, continuously maintained by a dedicated compressor. 
Between the CO$_2$ tank and the digital mass flow controller is placed a heater to avoid the presence of condensed CO$_2$ droplets that could change the flow dynamics and the instrument's response. For the same reason, a humidity valve is placed upstream of the air flow controller to collect any water droplets that may form in the compressed air line (Figure \ref{fig:CH1sketch}a).
Downstream the flow controllers, the three gas lines (air, CO$_2$, and C$_2$H$_6$) are intersected multiple times to homogenize the mixture before it enters the tranquillization chamber, which receives the flow through three separate inlets. The tranquillization chamber (Figure \ref{fig:CH1sketch}a) has a volume of 
$0.04$ m$^3$, and ends with a set of flow straighteners consisting of a 20 cm-long array of parallel channels, each with a circular cross-section of 6mm in diameter. Downstream of the flow straighteners, a vertically movable gate is installed to control the inlet current depth, which can be set between 0 and 10 cm. It is possible to block the passage of the mixture in the upper part of the flow straightener section. In this way, the effective height of the flow straightener section is set equal to the chosen inlet height.
At the end of the channel, the current flow is collected through a weir into a recovery tank, from which it is sucked and expelled from the building through a mechanical ventilation system. It was verified that this suction does not affect the current dynamics within the test section.  
Measurement probes (Section \ref{sec:CH1measurements}) are mounted on a mobile frame and move jointly. The wall-normal displacement of the measurement point is controlled digitally, while the horizontal and span-wise displacements are manual. The entire channel can be tilted, together with the probe displacement system, by up to 4 degrees.
The technical solutions adopted for the realization of the heated wall (Figure \ref{fig:CH1sketch}b) are reported in the following Section \ref{subsec:CH1heatedwall}.

\subsection{Heating wall}\label{subsec:CH1heatedwall}
The first 4 meters of the channel bottom wall can be warmed.  The heating elements (Vulcanic S.a.S.) consist of an electric resistance arranged in a serpentine pattern and embedded in a silicone matrix, resulting in a heating fabric with a thickness of 2.5 mm. The spacing of the serpentine pattern formed by the resistive wire is 7 mm. 
Eight fabrics, each with a surface of $497\times304$ mm$^2$, are positioned adjacent to one another to cover the first 4 meters of the channel. The temperature of the resistances inside the silicon, $T_{\Omega}$, is measured with a Pt$_{100}$ thermocouple.
The resistance of each of the wires is $54.5 \pm 0.5$ $\Omega$ at $T_{\Omega}=\SI{15}{\degreeCelsius}$, and increases linearly to $56 \pm 0.5$ $\Omega$ when the wire reaches its maximum operating temperature, i.e.  $T_{\Omega}= \SI{90}{\degreeCelsius}$.
The eight resistances are powered in parallel by a voltage controller, assuring a voltage $\Delta V$, and a total supplied heat flux per unit area $\varphi_{tot}$. 
Only a portion of $\varphi_{tot}$, denoted as $\varphi$, is effectively transferred by convection to the gravity current.
To ensure homogeneous heat transfer to the flow, a 1-cm-thick aluminium plate, with a thermal conductivity $\lambda_{Al}=137$ W/mK, is placed above the heating elements (Figure \ref{fig:CH1sketch}b). The upper surface of the aluminium plate is matte black painted to reduce reflections of the LDV laser beams. To minimize the heat losses below the heating elements, the following insulating layers are arranged: 2 cm of fiberglass ($\lambda_{fg}=0.03$ W/mK), 2.5 cm of vermiculite ($\lambda_{v}=0.09$ W/mK), and 3 cm of composite wood ($\lambda_w=0.18$ W/mK) (Figure \ref{fig:CH1sketch}b).  The same insulating sandwich constitutes the lower surface of the tranquilization chamber and the final 30 cm of the channel’s bottom wall, which are not heated.\\
Above the aluminium plate, three black matt-painted heat flux sensors are positioned on the channel centre-line at $x = 1$m, $2$m, and $3$m from the inlet, measuring the total heat flux (convective and radiative) exiting the aluminium plate, $\varphi_{meas}$. At the same locations, three thermocouples record the wall temperature, $T_w$. The variation of $\varphi_{meas}$ between the three measurement positions is below the $10\%$ and does not show a trend with the streamwise position, indicating a reasonably uniform heating distribution along the channel. The average between the 3 measurements of $\varphi_{meas}$ is considered.
To obtain a complete picture of the heat fluxes inside the channel bottom wall, the heat flux lost across the insulating materials below the heating elements is also measured. Thermocouples are embedded at three vertical locations at the interfaces between the different insulation materials (Figure \ref{fig:CH1sketch}b), enabling the estimate of the heat losses with Fourier's law for conduction \citep{cengel_2010}:
$\varphi_{iso} = \lambda \Delta T_{iso}/\Delta z_{iso}$, where $\lambda$ is the conductivity of the material considered, $\Delta z_{iso}$ is its thickness, and $\Delta T_{iso}$ is the temperature difference between its upper and lower surface. 
This estimate is conducted at four different streamwise positions. For all the tested conditions, the measured heat losses downward into the insulating material represent between 3\% and 5\% of  $\varphi_{tot}$.
Under steady-state temperature conditions, the difference between the total heat flux and the sum of $\varphi_{meas}$ and estimated losses in the insulating material,
$\varphi_{tot} - (\varphi_{meas} + \varphi_{iso})$,
is always positive and remains below 17\% of \(\varphi_{tot}\) (see Section \ref{sec:convection}). This residual is likely associated with a lateral heat flux escaping through the side boundaries of the wall.
An important observation is that the total heat flux measured by the flux sensors includes a radiation component, $\varphi_r$, exchanged between the aluminum channel floor and the room walls, therefore not contributing to the flux exchanged by convection, $\varphi=\varphi_{meas}-\varphi_r$. 
The flux $\varphi_r$  is estimated using the Stefan–Boltzmann law, $\varphi_r=\sigma \varepsilon (T_w^4 -T_0^4)$, where $\sigma$ is the Stefan-Boltzmann constant, $\varepsilon$ is the emissivity, and $T_0$ is the room temperature far from the wall.  The emissivity is assumed equal to 0.95, given that the aluminum wall has a matt black surface \citep{cengel_2010}. The portion of heat transferred by radiation can account for up to 65$\%$ of $\varphi_{meas}$. \cite{AhammadBasha2012} found a similar percentage of radiation contribution to the total heat transfer for mixed convection over horizontal highly emissive surfaces. In our analysis, an uncertainty of 0.05 is assumed for the emissivity value, which can translate up to an approximate 10\% uncertainty on the estimated convective heat flux $\varphi$. 
The value $\varepsilon = 0.95$ was selected as it provides the best agreement between the estimated convective heat fluxes and the excess of enthalpy and buoyancy fluxes directly measured in the current (Section \ref{sec:CH1results}).

Combining the uncertainty on the floor emissivity value and the variability among the $\varphi_{\mathrm{meas}}$ measurements at different $x$-positions, we estimate a total 15\% uncertainty for the spatially-homogeneous convective heat flux $\varphi$.

\subsection{Heat exchange characterization}\label{sec:convection}
Heat fluxes and wall temperature measurements, obtained as described in Section \ref{subsec:CH1heatedwall}, together with ambient temperature data, are employed in this Section to characterize the convection intensity between the heated wall and the overlying fluid, for different values of total power supplied to the heating fabrics. For comparison, the case of free convection -- where no imposed flow is present -- is also analyzed. The condition in which the gravity current is injected will be hereafter referred to as the mixed convection condition. The mixed convection results are obtained by imposing the same inlet conditions as those described in detail in Section \ref{sec:CH1inlet} but with the channel not tilted.  
A first notable observation is that, for all the tested conditions, the three wall temperature measurements, taken at the wall centre-line at three different distances from the source (Section \ref{subsec:CH1heatedwall}), differ by less than 2$\%$ (considering values in Celsius) and show no systematic variation along the streamwise direction. Therefore, a horizontally homogeneous wall temperature, $T_w$, defined as the average of the three measurements, will be considered in the following.

\begin{figure}[!h]
\centering
\includegraphics[width=0.9\textwidth]
{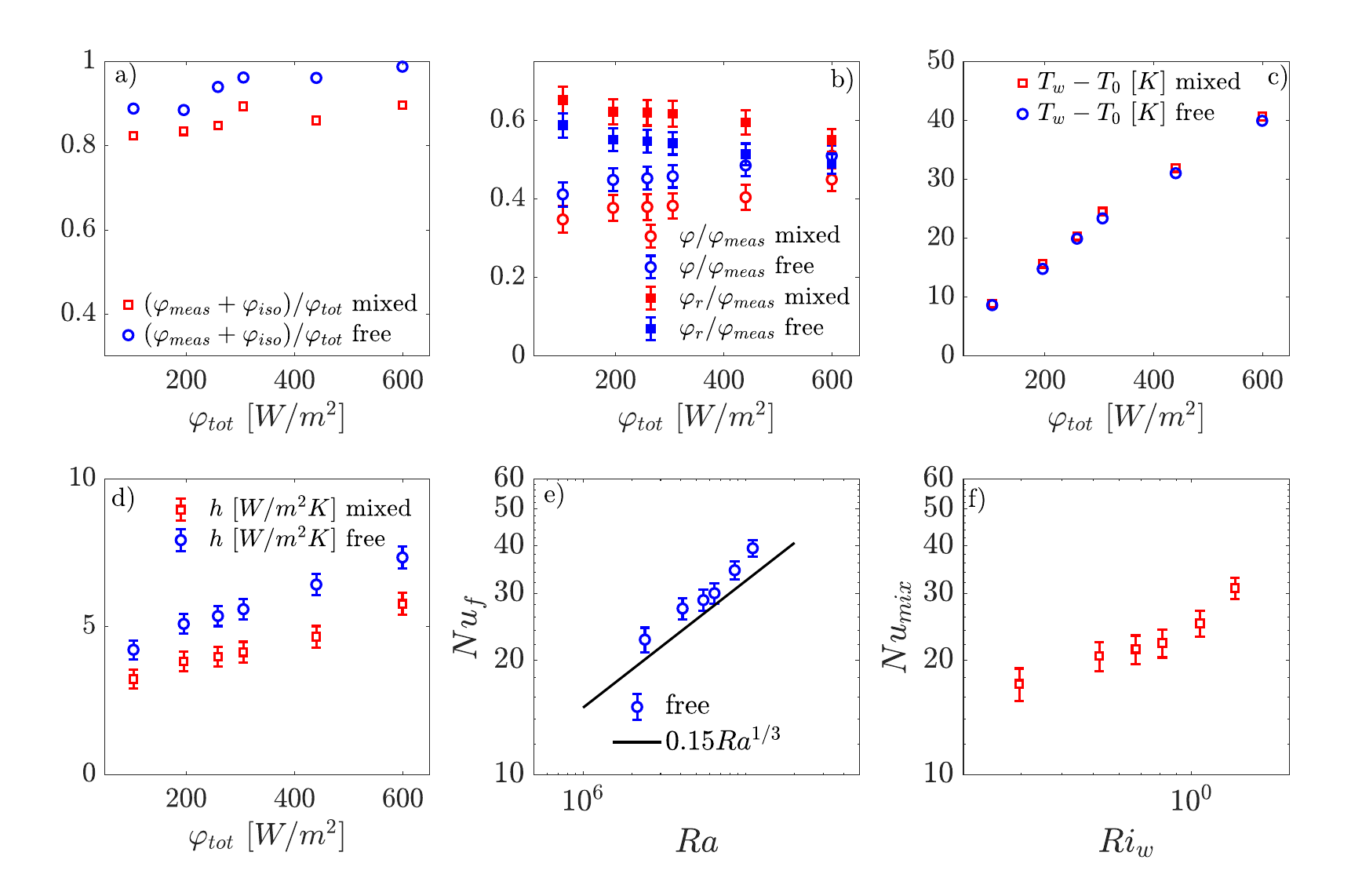}
\caption{Characterization of convection above the heating surface.  (b) Partition between the heat flux transferred by radiation and that transferred by convection, for both the free and mixed convection. (c) Temperature difference between the wall and the ambient.  
(d) Convective heat transfer coefficient, $h$.  Panels (e) and (f) show the Nusselt numbers for free and mixed convection, respectively.}
\label{fig:convection}
\end{figure}

Figure \ref{fig:convection}a shows the ratio between the sum of the heat flux normal to the aluminum wall ($\varphi_{meas}$) and the downward conductive flux lost in the isolant ($\varphi_{iso}$), and the total flux per unit area produced by the electrical resistance ($\varphi_{tot}$). Values smaller than one indicate that part of $\varphi_{{tot}}$ is lost due to lateral conductive heat transfer through the aluminum and insulating walls. The lateral loss is higher in the mixed convection case with respect to the free convection.
The portions of heat exchanged via radiation, $\varphi_r/\varphi_{meas}$, and convection, $\varphi/\varphi_{meas}$, for both natural and mixed convection cases are presented in Figure \ref{fig:convection}b. The error bars show the range of possible results considering the emissivity $0.9<\varepsilon<1.0$, while the markers correspond to the values obtained for $\varepsilon=0.95$. The results indicate that, for increasing $\varphi_{tot}$, the portion of heat transferred by convection increases for both free and mixed convection, and consequently, the fraction exchanged via radiation is reduced.
Moreover, for a fixed $\varphi_{tot}$, the convective heat transfer is greater in the case of free convection than in the case of mixed convection.   This observation is associated with a slightly higher wall temperature difference $T_w-T_0$ in the case with imposed flow compared with the no-flow case (Figure \ref{fig:convection}c), and with a lower convective heat transfer coefficient, $
h = \varphi/(T_w - T_0) $, in the mixed convection condition than in free convection (Figure \ref{fig:convection}d).
These results, although contrary to initial expectations, can be justified by the following arguments.
The presence of the gravity current can reduce the convective heat transfer compared with the no-current case, as the shallow horizontal flow is capable of disrupting the natural thermal plumes whose vertical extent would otherwise be much greater and reach the ceiling of the room. Owing to its relatively low Reynolds number (Section \ref{sec:CH1inlet}) and reduced vertical extent with respect to the free-convective thermal plumes,  the gravity current-related mixed convection may be less effective than that driven by free thermal plumes. The reduced efficiency of mixed convection relative to free convection is consistent with Figure\ref{fig:convection}a, which indicates that a decrease in upward convective heat transfer in the mixed condition is accompanied by an increase in lateral conductive heat transfer through the aluminium and insulation walls. 

Figures~\ref{fig:convection}d and ~\ref{fig:convection}e  show the Nusselt numbers characterizing the intensity of free ($Nu_f$), and mixed ($Nu_{mix}$) convections,  respectively.
The Nusselt numbers are defined as
\begin{equation}
    Nu_*=\frac{hL}{\alpha_{0}\rho_0 c_{p,0}},
\end{equation}
where  `*' can be substituted by `f' or `mix', $\alpha_0$, $\rho_0$ and $c_{p,0}$ are the thermal diffusivity, density and heat mass capacity at constant pressure of the air far from the wall (see Section \ref{subsec:CH1mixtureproperties})  and the characteristic length is defined as $L=A/p$, where $A$ is the bottom wall area and $p$ its perimeter. The Rayleigh number, characterizing the free convection regime, is defined as
\begin{equation}
   Ra = \frac{g (T_w-T_0) L^3}{\nu_{0} \alpha_{0}T_0},
\end{equation}
where $\nu_{0}$ is the air kinematic viscosity in standard conditions. 
Figure~\ref{fig:convection}d presents the values of $Nu_f$ as a function of $Ra$, along with a reference empirical law for natural convection over a flat plate heated from below and without lateral losses, $Nu_f=0.15Ra^{1/3}$ \citep{cengel_2010}.
When the current is injected, a characteristic Richardson number, characterizing the ratio between the intensity of free convection and the forced one, can be defined as
\begin{equation}
    Ri_w=g(T_w -T_0) L/T_0u_s^2,
\end{equation}
 where $u_s$ is the bulk inlet velocity.   Figure \ref{fig:convection}e presents $Nu_{mix}$ as a function of $Ri_w$. As wall heating increases, $Ri_w$ rises due to the enhanced influence of natural convection, which concurrently leads to an increase in the Nusselt number. This observation, namely the dependence of the $Nu_{mix}$ on $Ri_w$, demonstrates that even when the current is present, the flow cannot be classified as purely forced convection. The latter would, in fact, imply that the Nusselt number is independent of $Ri_w$ and that the intensity of convective exchange depends solely on the inlet conditions of the current, in particular on its Reynolds number. It should also be noted that the choice of length scale adopted in the definition of $Ri_w$ is arbitrary, and alternative definitions could be based on a characteristic height associated with the current.

\section{Measurement Techniques}\label{sec:CH1measurements}

\subsection{LDV}\label{subsec:CH1LDV}
The horizontal (i.e., along $x$) and vertical (i.e., along $z$) velocity components with respect to the channel wall are measured using a Laser Doppler Velocimeter (LDV) Argon class IV, equipped with a 5W power laser. It produces two blue and two green beams in perpendicular planes with wavelength $\Lambda_{blue}=488$ nm and $\Lambda_{green}=514.5$ nm, respectively.   The front lens has a focal length of $400$ mm, and the beams have a diameter of $0.1$ mm. Backscatter detection is used. An accurate evaluation of velocity statistics requires seeding both the flow introduced through the channel inlet (internal seeding) and the surrounding ambient air (external seeding). 
The internal seeding is obtained with micronic oil droplets, injected into the mixture upstream of the tranquilization chamber using an atomizer that generates polydisperse droplets with a typical diameter of 1 $\mu$m \citep{Marro2020}.  The external seeding is obtained
with a fog generator using the SAFEX-Inside-Nebelfluid Dräger Spezial W fluid, emitting droplets in the range 0.5-2 $\mu$m \citep{DelPonte2024}. The fog generator is positioned on the lateral side of the channel to ensure proper homogenization of the fog before it enters the interior of the channel. 
The double-seeding system allows for maintaining the LDV data rate between 500 Hz and 1500 Hz. For each measurement point, at least 100000 values are collected. We verified that these acquisition parameters guarantee a good convergence of the first-
and second-order velocity statistics.  

\subsection{FID}\label{subsec:CH1FID}
\begin{figure}
\centering
\includegraphics[width=1\textwidth]{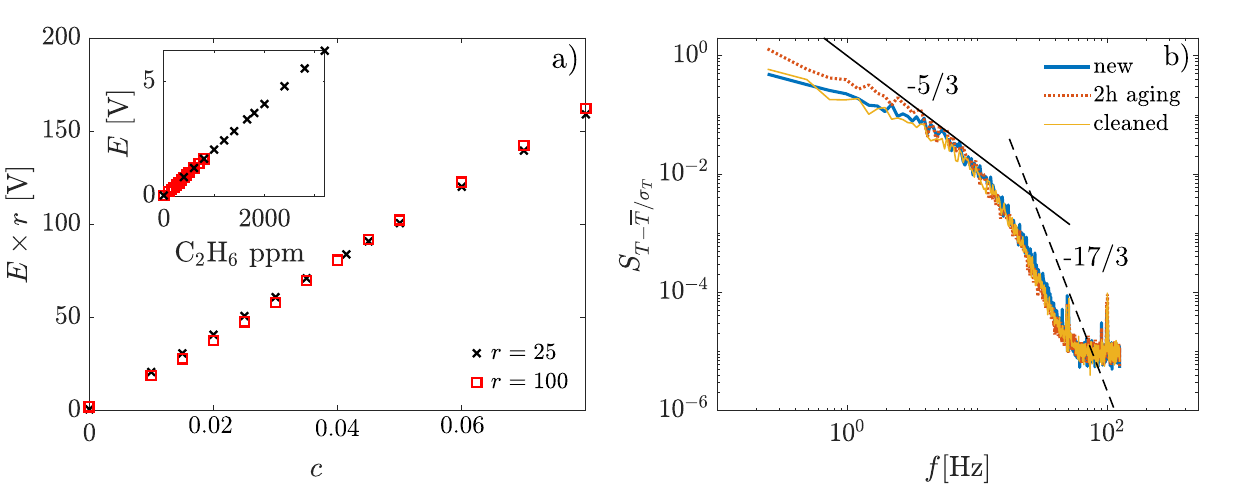}
\caption{(a) FID voltage output E times $r$ as a function of CO$_2$ volume fraction, $c$, for $r=25$ and $r=100$. The inset shows the FID voltage output as a function of the C$_2$H$_6$ ppm in the mixture. (b) Example of temperature spectrum obtained with the CCA with $d=1\mu$m. The three curves show the spectrum before and after the exposure to the seeding and after the cleaning with a jet of pure air.}
\label{fig:fidcca}
\end{figure}
Measurements of CO$_2$ concentration are performed using a Flame Ionization Detector (FID), with ethane serving as tracer \citep{Vidali2022}. This procedure allows measuring the CO$_2$ volume fraction $c$ in the mixture. A mixture sample is sucked into a 0.3 m  long tube with a diameter of 0.125 mm and burned in the combustion chamber. The resulting ions are collected at the cathode, and the instrument measures the potential difference generated by the ionization current, which is proportional to the ethane molecules in the sample \citep{Fackrell1980}. The effective FID frequency response is approximately 800 Hz \citep{Marro2020}.  In our experiments, the presence of substances different from the tracer, such as CO$_2$ or seeding droplets, may interfere with the combustion process and alter the instrument response.  
The FID calibration procedure in the presence of carbon dioxide, described in detail in \cite{Vidali2022}, is adopted. The estimate of the CO$_2$ volume fraction from FID measurements of ethane volume fraction $c_{\text{C}_2\text{H}_6}$, relies on the assumption that the ratio $ r=c/c_{\text{C}_2\text{H}_6}$ imposed at the inlet remains constant throughout the mixing process. This assumption is justified by the fact that the molecular diffusion coefficients of CO$_2$ and C$_2$H$_6$ in air are comparable, \( D_{\mathrm{CO_2}} = 1.61 \times 10^{-5} \, \mathrm{m^2/s} \) and \( D_{\mathrm{C_2H_6}} = 1.48 \times 10^{-5} \, \mathrm{m^2/s} \) at 20\,$^\circ$C \citep{Pritchard1982}, and that the gases are well blended at the inlet.   \cite{Vidali2022} found that, for gas mixtures with a CO$_2$ volume fraction $c>0.58$, the calibration curve (between the ethane volume fraction and the FID output voltage) is no longer bijective, thereby preventing quantitative measurements. However, the present study will focus on Boussinesq currents with an inlet CO$_2$ volume fraction of $c_s=0.07$, for which the calibration remains bijective and linear.  
To avoid FID saturation and to exploit its dynamic range (0-10$V$), $r$ values of 25 and 100 are used in the experiments.  The calibration curves for $r=25$ and $r=100$ are shown in Figure \ref{fig:fidcca}a. The FID calibration was performed at least twice per day and was repeated if the flame temperature showed variations of more than 10\,°C  from the value recorded at the beginning of the experiment. \\
The effect of seeding droplets on FID measurements has been analyzed in detail by \cite{Marro2020}. They showed that seeding droplets induce a slightly higher uncertainty in scalar statistics that, however, does not exceed 5\% and does not compromise the reliability of turbulent mass flux estimates of the LDV-FID system.\\
In the remainder of the work, the term CO$_2$ \textit{concentration} will always refer to the volume fraction (which is equal to the molar fraction) of CO$_2$ in the mixture, unless stated otherwise.

\subsection{Cold Wire}\label{subsec:CH1CCA}
Temperature measurements are performed by means of a cold wire probe, consisting of a thin Platinum-Rhodium10$\%$ wire operated with a very low constant current and used as a resistance temperature detector \citep{Fulachier1978}. Cold wire-probes with two different diameter, $d=1\mu$m and $d=2.5\mu$m, are used.  Specific considerations must be taken into account in order to employ this measurement technique in our experiment, since the flow conditions differ from those of pure air, where cold-wires are typically used. The first consideration is that the presence of CO$_{2}$ and C$_2$H$_6$ in the mixture may affect the cold-wire response. 
\cite{Paranthoen1982} have shown that, when the cold-wire is used in mixtures with thermal diffusivities significantly higher than that of air, the formation of a thick temperature boundary layer around the wire supports can alter the CCA response with respect to the pure-air condition, even at low frequencies. However, this effect becomes negligible when the gas has a thermal diffusivity similar to that of air, such as argon \citep{Paranthoen1982} or air–helium mixtures with low helium concentrations \citep{Hewes_2022}. For our experiments, since the CO$_2$ and C$_2$H$_6$ thermal diffusivities at 25$^\circ$C  and atmospheric pressure, $\alpha_{\text{CO}_2}=1.09\times 10^{-5}$ m$^2/$s and $\alpha_{\text{C}_2\text{H}_6}=0.9\times10^{-5}$ m$^2/$s, are of the same order of magnitude of that of air, $\alpha_{0}=2.12\times10^{-5}$ m$^2/$s \citep{cengel_2010}, and since the maximum volume fraction of CO$_2$ (C$_2$H$_6$) at the inlet is $c_s=0.07$ ($0.07/r$), we assume that the response of the cold-wire probe is insensitive to the composition of the gas mixture, allowing CCA calibration to be performed in pure air. 
A second important aspect to consider is the possible deposition of oil droplets on the cold wire. \cite{Weiss2005} investigated this phenomenon in detail and showed that the transfer function of the cold-wire probe is progressively attenuated at medium and high frequencies as the aging time of the sensor increases. However, they demonstrated that the attenuation primarily affects frequencies higher than the inverse of the time constant associated with the deposited droplets. After two hours of exposure to the seeded mixture, droplets with diameters up to $25\ \mu$m may form on the wire, corresponding to a time constant of approximately $3 \times 10^{-3}$ s, i.e., a frequency of about 333 Hz \citep{Weiss2005}. Adopting the isotropic approximation and Taylor’s hypotheses
of frozen turbulence, the turbulent dissipation rate $\epsilon$ can be computed from a point-wise time-resolved signal as $\epsilon=(15 \nu /\overline {u}^2) \overline{(\text{d}u/\text{d}t)^2}$ \citep{Hinze1959}, where $\nu$ is the kinematic viscosity. 
This allows an estimate of the Kolmogorov time scale $\tau_\eta \approx 3 \times 10^{-3}$ s  \citep{Pope_2000} in the region where $\overline{u} > 0.3u_s$. This implies that, according to \cite{Weiss2005}, the cold-wire transfer function is only attenuated at frequencies higher than those related to the Kolmogorov time scale in our measurement. Therefore, in the present study, the effect of wire aging is negligible for the estimate of first- and second-order statistics. This conclusion is supported by the comparison shown in Figure \ref{fig:fidcca}b, where temperature spectra obtained at $x=1900$mm and $z=35$mm in the $\Lambda_2$ experiment (see Section \ref{sec:CH1inlet}) using the $1\mu$m new wire and the same wire aged for two hours in the seeded flow are reported. To remove the bigger oil droplets, at the beginning of each measurement day, the CCA is cleaned with a jet of pure air. Beyond the $-5/3$ region in the inertial--convective range, a second scaling regime can be observed, which aligns well with the $-17/3$ law predicted by \cite{Batchelor1959} and reported by \cite{Heist1998} in a stable boundary layer, and is here also identified in a gravity current advancing over a heating wall. Similar measurements performed with a $2.5\,\mu\text{m}$ diameter wire yield practically identical results.

\begin{figure}
\centering
\includegraphics[width=1\textwidth]
{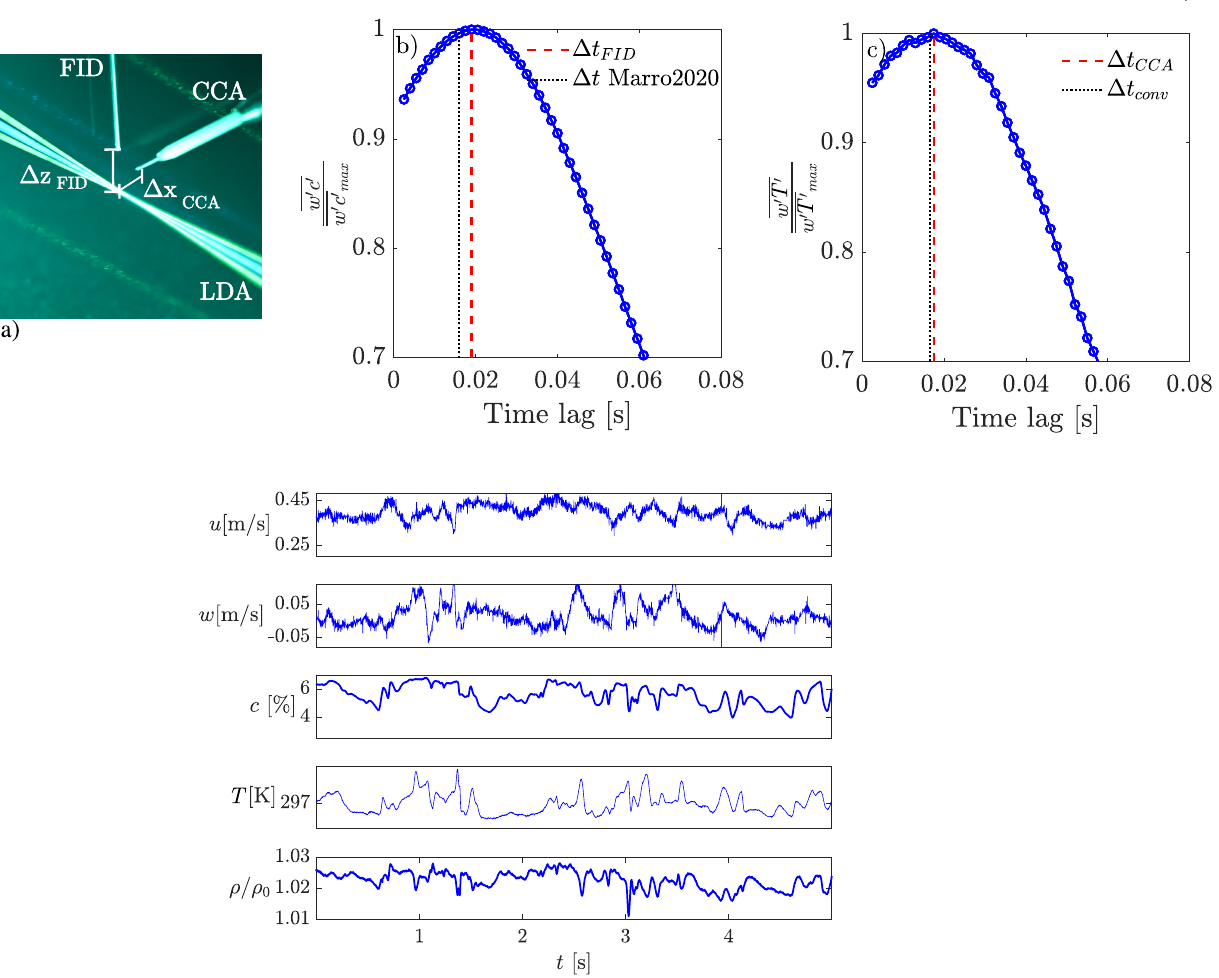}
\caption{(a) Picture of the experimental probes' position. Panels (b) and (c) show the values of the cross-correlations $\overline{w'c'}$ and $\overline{w'T'}$ as a function of the time lag of the concentration and temperature signals with respect to the velocity one, respectively. Panel (d) shows an example of the longitudinal velocity, vertical velocity, CO$_2$ concentration, temperature, and density signals that are simultaneously recorded during the experiment.}
\label{fig:signals}
\end{figure}
\subsection{Cross-correlation and density signal reconstruction}\label{subsec:CH1signals}

The position of the probes relative to each other is shown in Figure \ref{fig:signals}(a). This was chosen as a compromise between the requirement that the distance between the probes is sufficiently small that they measure the same fluid volume and sufficiently large that the presence of one probe should not interfere with the other. The FID is an intrusive method performing measurement by aspirating the fluid to a flame chamber, which implies the slight local modification of the flow field. Following  \cite{Marro2020}, the FID sampling probe is therefore positioned $4$ mm above the LDV measurement volume ($\Delta x_{FID}=0$mm, $\Delta z_{FID}=4$mm).  Conversely, the cold-wire probe is located at the same vertical height of the LDV measurement volume but $5$ mm downstream of it ($\Delta x_{CCA}=5$mm, $\Delta z_{CCA}=0$mm). 
In the region where $ \overline{u} > 0.3 u_s$, the Kolmogorov length scale $\eta=(\nu^3/\epsilon)^{1/4}$ lies between $0.3\,\mathrm{mm}$ and $0.5\,\mathrm{mm}$, therefore $\Delta z_{FID}$ correspond to about 8-13$\eta$ and $\Delta x_{CCA}$ to about 10-15$\eta$.  The spatial separation $ \Delta x_{CCA} \approx 10-15 \eta $ used in this study is slightly larger than that employed in previous combined LDV-CCA measurements of turbulent heat fluxes (e.g. \cite{Wardana1995,Heist1998,Pietri2000, Darisse2013}), and can cause up to a 5-10\% underestimation of the velocity-temperature correlation coefficient, if the signals are not shifted in time \citep{Heist1998}. However, this error can be reduced by shifting the temperature signal in time compared to the velocity one, with a method that will be explained later in this section.\\
The protocol adopted to estimate the turbulent fluxes of CO$_2$ concentration, $\overline{u'c'}$ and $\overline{w'c'}$, and turbulent heat fluxes, $\overline{u'T'}$ and $\overline{w'T'}$, is the same as that presented in \cite{Marro2020} for LDV-FID system and is here extended to include the acquisition and processing of the temperature signal.  The acquisition system (LDV-FID-CCA) is set in order to be governed by the LDV, i.e., the velocity, concentration, and temperature signals are simultaneously collected when the seeding particles cross the LDV sample volume. Therefore, the sampling frequency of temperature and concentration is also irregular.
To minimize the errors in the turbulent flux estimates, both the FID and CCA signals are shifted by a time lag $\Delta t_{FID}$ and $\Delta t_{CCA}$ with respect to the LDV signal. 
The cross-correlations are then obtained with the sample-and-hold reconstruction and resampling method \citep{Marro2020}, which consists of i) resampling the temperature and concentration values on the temporal pattern of LDV data, i.e., on velocity signals, and ii) computing the cross-correlation with transit time weighting.
In general, $\Delta t_{FID}$ and $\Delta t_{CCA}$ are different. For each measurement point, different time shifts are tested, and the ones that maximize the $\overline{w'c'}$ and $\overline{w'T'}$ correlations are adopted for the concentration and temperature acquisitions, respectively. An example of the $\overline{w'c'}$ and $\overline{w'T'}$ values as a function of the time lag with respect to the velocity signals is presented in Figure \ref{fig:signals}(b,c). 
At this location, the maximum value of $\overline{w'c'}$ is obtained by delaying the concentration signal by $ \Delta t_{FID} = 18.5$ ms. This value is close to the 16.0\,ms time lag reported by \cite{Marro2020}, obtained using the same FID employed in the present study. That delay corresponds to the time required for the gas sample to travel through the 0.3\,m  long FID tube and reach the combustion chamber. The maximum of $\overline{w'T'}$ is obtained shifting the temperature signal by $ \Delta t_{CCA} = 18.5$ ms. This value can be predicted with reasonably good accuracy, considering the mean time $\Delta t _{conv}$ required for the fluid particle to travel the distance $\Delta x_{CCA}$, assuming a mean horizontal velocity equal to the one measured by the LDV. An analogous shifting technique of the CCA signal is adopted to compensate for convection time, for example, in \cite{Wardana1995}. 
\subsection{Mixture properties}\label{subsec:CH1mixtureproperties}
Relying on the time-shifted volume fraction and temperature signals, $c$ and $T$, it is possible to reconstruct the local properties of the mixture, like its molar mass, specific heat, and density. 
For this analysis, the contribution of ethane to the mixture properties is considered negligible. This assumption is justified by the very low concentration of ethane in the mixture, equal to $c/r$, and by the fact that its molar mass, $M_{\text{C}_2\text{H}_6}=30.07$ g/mol, is very close to that of pure air, $M_{\text{0}}=28.97$ g/mol.

The ratio between the molar mass of the air-CO$_2$ mixture, $M$, and $M_0$ can be written as a function of the CO$_2$ volume fraction in the mixture $c$ as:
\begin{align}
    \frac{M}{M_0} = c \biggl ( \frac{M_{\text{CO}_2}}{M_0} -1\biggr ) +1= c\chi_M +1, \label{eq:molarmasses}
\end{align}
where $M_{\text{CO}_2}=44.01$ g/mol, is the CO$_2$ molar mass and $\chi_M=0.519$ is a constant that depends only on $M_{\text{CO}_2}$ and $M_0$. The perfect gas law, that can be written for the  mixture or for the external pure air, is
\begin{equation}
\frac{\rho T}{ M } =\frac{\rho_0 T_0}{M_0}= \frac{P_0}{R}, 
\label{eq:CH1perfect_gas_law}
\end{equation}
where $P_0$ is the ambient pressure, $R$ is the ideal gas constant, 
and $\rho_0$ and $T_0$ are the pure air density and temperature outside the gravity current. Since we are dealing with low-Mach number flow, $P_0$ can be considered constant in the entire flow domain. Manipulating equations \ref{eq:molarmasses} and \ref{eq:CH1perfect_gas_law}, allows obtaining the instantaneous local value of the mixture density $\rho$ as a function of $c$ and $T$. An example of the time-resolved signals of horizontal velocity, vertical velocity, CO\textsubscript{2} volume fraction, temperature, and density obtained in one measurement point is shown in Figure \ref{fig:signals}(d).

In this work, the volumetric fraction of CO$_2$ does not exceed $c = 0.07$, and the temperature variations due to wall heating, $(T-T_0)/T_0=\Delta T/T_0$, are limited to approximately 10$\%$. The density variation induced by $c$ or $\Delta T$ is small compared to the reference density $\rho_0$ and therefore, the Boussinesq approximation can be considered valid to describe the flow dynamics. These observation allows us to obtain the relations shown in the following of this Section.
Assuming that the variations of temperature are small and manipulating  equations \ref{eq:molarmasses}  and \ref{eq:CH1perfect_gas_law}, the buoyancy $b=g(\rho-\rho_0)/\rho_0$ can be approximated as:
\begin{equation}
b \approx  g \biggl(\chi_M c - \frac{\Delta T}{T_0}\biggr).
\label{eq:CH1bapprox}
\end{equation}
The CO$_2$ mass fraction $\tilde{c}$ is written as a function of the volume fraction $c$, as:
\begin{align}
    \tilde{c}\DEFN c \frac{M_{\text{CO}_2}}{M}=c\frac{M_{\text{CO}_2}}{M_0}\frac{1}{1+c\chi_M}\approx c\frac{M_{\text{CO}_2}}{M_0}.\label{eq:massfraction}
\end{align}
Finally, the heat mass capacity at constant pressure of air and carbon dioxide is given, respectively, by: $c_{p,0} = f_0 R/M_0$, $c_{p,\text{CO}_2} = f_1 R/M_1$, with $f_0 \approx 7/2$, $f_1 \approx 9/2$ for a temperature close to 25°C. 
The specific heat mass capacity at constant pressure of the mixture is:
\begin{align}
   \frac{c_p}{c_{p,0}} =   \biggl( 1 + \frac{c_{p,\text{CO}_2 }-c_{p,0}}{c_{p,0}} \tilde{c} \bigg)= \bigg ( 1+ \chi_{c_p}c\bigg)\approx 1,\label{eq:cp}
\end{align}
where $\chi_{c_p}=\frac{c_{p,\text{CO}_2 }-c_{p,0}}{c_{p,0}}\frac{M_{\text{CO}_2}}{M_0}=-0.231$. Therefore, the $c_p$ coefficient in the mixture will be considered constant and equal to $c_{p,0}$ in the following.

\section{Experimental Conditions}\label{sec:CH1inlet}
\begin{figure}
\centering
\includegraphics[width=1\textwidth]{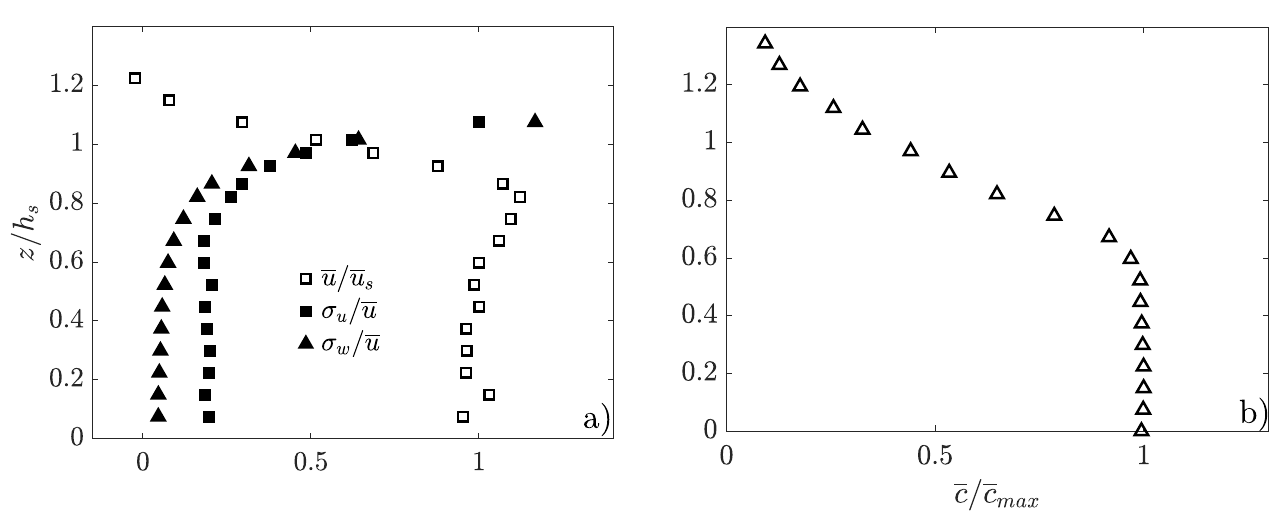}
\caption{Vertical profiles measured 1$h_s$ downstream the inlet: panel (a) shows the mean streamwise velocity and the horizontal and vertical turbulent intensity, panel (b) shows the mean CO$_2$ concentration.}
\label{fig:inlet}
\end{figure}
For all the experiments, the channel inclination is fixed at $\phi= 3$\,°. Although small, this slope is sufficient to maintain a supercritical regime along the entire channel \citep{Sequeiros2012}, thereby preventing the outlet condition from affecting the whole flow domain \citep{Haddad2022,Ungarish2023}. 
This condition will also hold for the heated currents, as will be demonstrated in the Section \ref{subsec:CH1integrals}.
Tilted channels are often employed in adiabatic gravity-current experiments (e.g. \cite{Parker1987, Krug2013, Odier2014, Martin_Negretti_Hopfinger_2019}), particularly to investigate currents approaching the normal condition, i.e., a region sufficiently far from the source where the bulk Richardson number attains a constant value. 
The inlet height is set as $h_s=6.7$ cm, and the imposed CO$_2$ volume fraction in the mixture at the inlet is $c_s=0.07$. This corresponds to a density ratio at the source $\rho_s/\rho_0$ equal to 1.036, where $\rho_0$ is the ambient density, and a source buoyancy $b_s=g(\rho_s -\rho_0)/\rho_0=0.36$ m/s$^2$, where $g$ is the gravitational acceleration.   The vertical profiles at $x=h_s$ of the mean streamwise velocity, $\overline{u}$, and the turbulence intensities $\sigma_u/\overline{u}$ and $\sigma_w/\overline{u}$, where $\sigma_u$ and $\sigma_w$ are the wall-parallel and wall-normal velocity standard deviations, are shown in Figure \ref{fig:inlet}a.  Due to limitations related to the displacement of the measurement position, $x=h_s$ is the closest distance downstream of the inlet at which measurements can be performed. The velocity profile is nearly top-hat, with variations smaller than $\pm10\%$ for $z < 0.8 h_s$. In this region, the streamwise turbulence intensity is in the range  $0.18<\sigma_u/\overline{u}<0.20$. For $z/h_s > 0.8$, a mixing zone with almost constant $\partial \overline{u}/\partial z$ and characterized by a peak in $\sigma_u/\overline{u}$ and  $\sigma_w/\overline{u}$ is present. The inlet bulk velocity, obtained by averaging the vertical velocity profile, is $u_s=0.36$m/s. Figure \ref{fig:inlet}b, shows the concentration profile 1$h_s$ downstream of the inlet.
The Richardson number at the source $Ri_s=(\cos \phi) b_s h_s/u_s^2$ is equal to 0.174, and the Reynolds number is $Re_s=u_sh_s/\nu_s$ is 1710, where $\nu_s=1.45\times 10^{-5}$ m$^2$/s is the mixture viscosity at the source.
With these inlet conditions, the experiments are repeated for three different heating intensities: the unheated case (with an imposed voltage at the resistances equal to $\Delta V_0=0$V), an intermediate heated case (with $\Delta V_1=29$V), and a high heated case (with $\Delta V_2=39$V). Once the heating is turned on, thermal steady-state is awaited before starting the experiments. The latter is reached when the wall temperatures no longer exhibit any temporal variation. This stabilization period lasts approximately 3 hours. During the first 2 hours, wall-heating is applied without any flow; during the final 60 minutes, the flow is also activated. 
At the steady state, the three voltages, $\Delta V_0$, $\Delta V_1$, and $\Delta V_2$,  correspond to a convective heat flux per unit area transferred to the current of $\varphi_0=0$ W$/$m$^2$, $\varphi_1=30 \pm 4$ W$/$m$^2$, and $\varphi_2=50 \pm 7$ W$/$m$^2$, respectively.  A non-dimensional number, $\Lambda_s$, that quantifies the intensity of the current heating can be defined as:
\begin{equation}
    \Lambda_s=\frac{g\varphi}{\rho_0 c_{p,0} T_0 b_s u_s}.
    \label{eq:ch1lambda}
\end{equation}
This parameter is obtained considering the ratio between the wall-normal buoyancy flux per unit area, induced by the heating $\varphi$, $\varphi g/ \rho_0 T_0  c_{p,0}$ \citep{Linden1999}, and the wall-tangent buoyancy flux per unit area that generates the currents, $b_su_s$. 
The physical meaning of this parameter is further discussed in Section \ref{subsec:CH1integrals}.
The $\Lambda_s$ values corresponding to the three heating intensities, $\varphi_0$, $\varphi_1$, and  $\varphi_2$ are $\Lambda_0=0$, $\Lambda_1=0.0065$, and $\Lambda_2=0.0110$, respectively.
These three heating conditions result in the longitudinal integral buoyancy flux at the end of the channel being equal to approximately $100\%$, $60\%$, and $35\%$ of its value at the source, respectively (see Section \ref{subsec:CH1integrals}).

\section{Results}\label{sec:CH1results}
The results presented in this Section are obtained by performing measurements in the centre-line of the channel, under the assumption that, sufficiently far from the lateral walls, they are independent of the span-wise position $y$. This will be verified by integral balances presented in Section \ref{subsec:CH1integrals}. 
\subsection{Flow Visualization}
\begin{figure}
\centering
\includegraphics[width=0.5\textwidth]{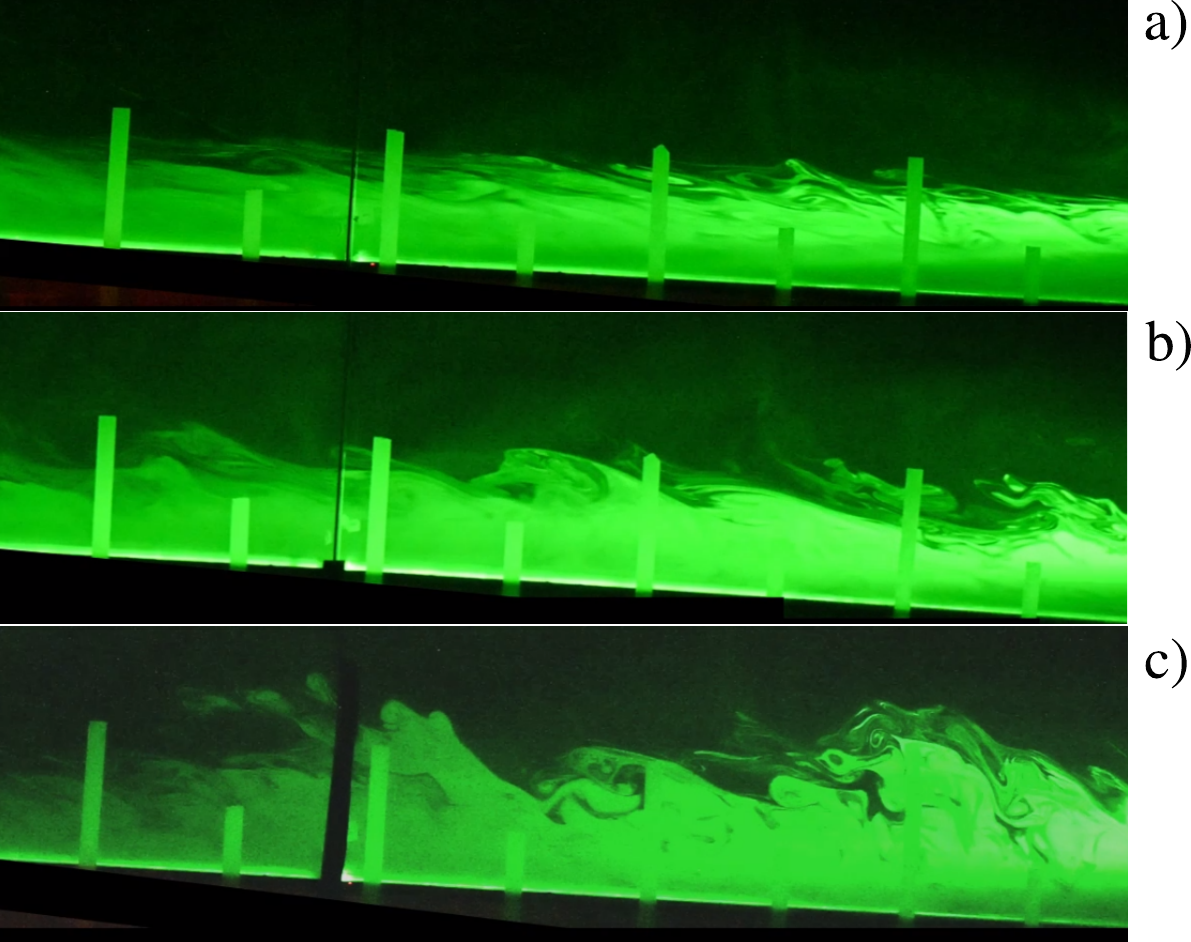}
\caption{Visualizations of the statistically-steady currents, in the zone $2$ m $<x<3.3$ m of the $\Lambda_0$(a), $\Lambda_1$(b), and $\Lambda_2$(c) currents.}
\label{fig:CH1visual}
\end{figure}
A phenomenological description of the flow can be derived from the flow visualizations shown in Figure \ref{fig:CH1visual}. The visualizations are obtained through laser tomography seeding the injected mixture with olive-oil droplets and using a green laser sheet at the channel's symmetry plane. The Figure shows the portion of the channel for $2$m$<x<3.3$m. As expected, heating alters the flow structure, significantly enhancing turbulence intensity and increasing the maximum height reached by the current mixture. The intensity and the vertical extent of the instabilities increase with $\Lambda_s$. In particular, vertical motions induced by thermals generated by heating are superimposed on the classical Kelvin--Helmholtz instability structures that arise in the adiabatic case, making the interface between the current and the ambient fluid considerably more chaotic.

\subsection{First-order statistics}\label{subsec:CH1firstorder} 
\begin{figure}[!ht]
\centering
\includegraphics[width=0.9\textwidth]{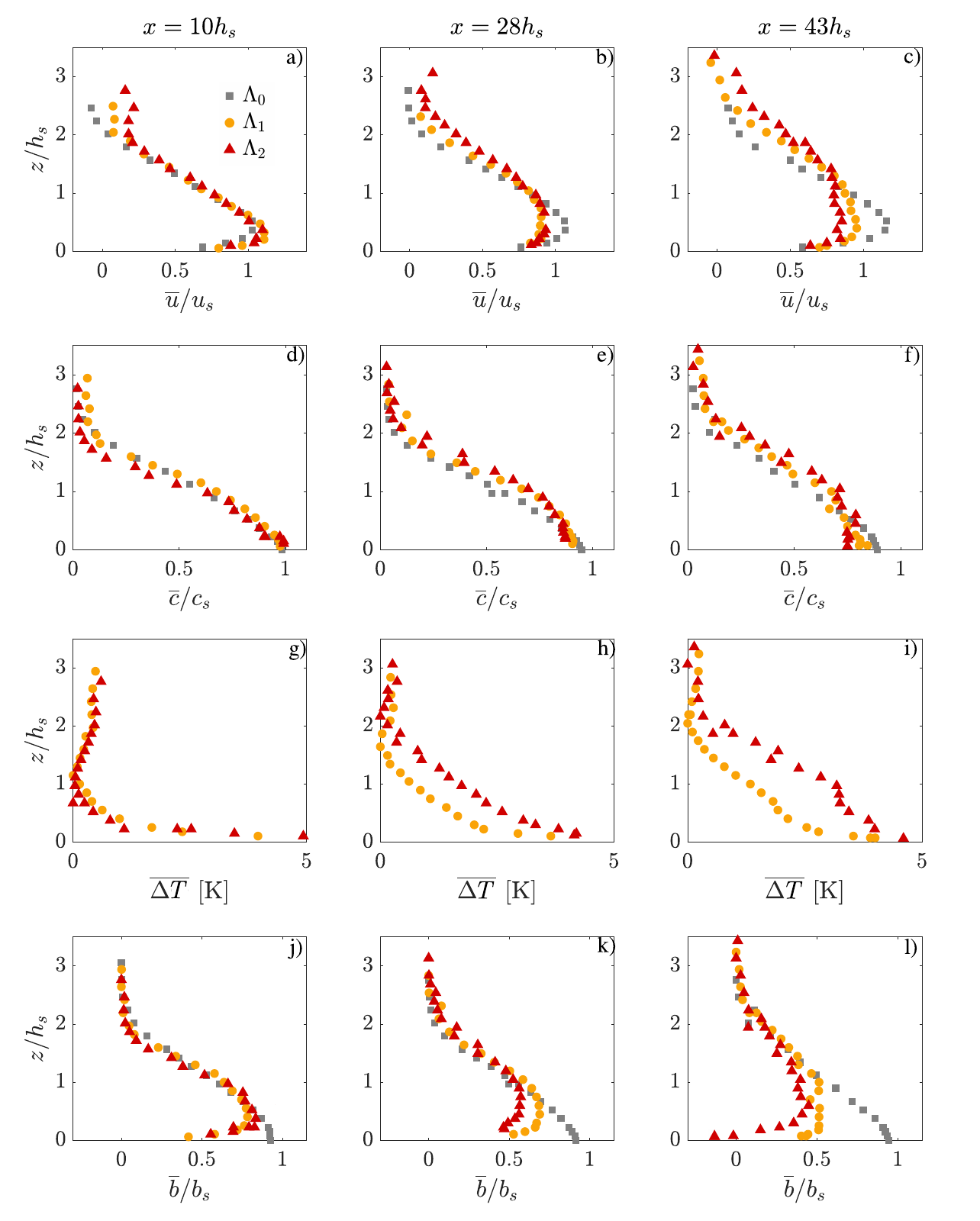}
\caption{Vertical profiles of mean horizontal velocity (a-c) and concentration (d-f) scaled with the inlet values of $u_s$ and $c_s$, respectively. Panels (g-i) show the vertical profiles of the mean temperature difference with respect to $T_0$ and (j-l) the vertical profiles of the mean buoyancy. The profiles are shown at three different distances from the inlet: 10$h_s$, 28$h_s$, and 43$h_s$. }
\label{fig:profilemean}
\end{figure}

In this Section, the vertical and horizontal profiles (with respect to the channel floor) of the first-order statistics are presented. The overline indicates the temporal averaging.
Figure \ref{fig:profilemean} shows the mean vertical profiles of horizontal velocity (Figure \ref{fig:profilemean}a-c) and CO$_2$ concentration  (Figure \ref{fig:profilemean}d-f), scaled with their source value, $u_s$ and $c_s$, respectively, temperature differences with respect to the ambient temperature, $T_0$ (Figure \ref{fig:profilemean}g-i), and buoyancy scaled with $b_s$ (Figure \ref{fig:profilemean}j-l). The profiles are shown at three distances from the inlet: $10h_s$, $28h_s$, and $43h_s$.  For all the quantities, only small variations with heating are observed at 10$h_s$. The differences between the heating levels become more pronounced for larger distances from the source. 
As for the non-heated case \citep{Kneller1999, Odier2014, Horsley2018}, even in the heated experiments, the shape of the velocity profiles (Figure \ref{fig:profilemean}a-c) can be approximately divided into three layers consisting of a near-wall boundary layer, an intermediate region with approximately uniform velocity, and a third region characterized by an almost-constant velocity gradient $\partial \overline{u}/\partial z$. The latter will be referred to as the `mixing region' in the remainder of the paper.  The vertical extent of the intermediate region increases both with the heating intensity and with distance from the source, and the maximum velocity is reduced. At $x=43h_s$, the maximum velocity decreases by 26$\%$ for $\Lambda_2$ and 17$\%$ for $\Lambda_1$ compared to the unheated case (Figure \ref{fig:profilemean}c).  
The effects of wall heating are clearly observable also in the concentration profiles (Figure \ref{fig:profilemean}d-f). Heating leads to a reduction of the concentration at the wall and a decrease in the norm of the near-wall gradient $\partial \overline{c}/\partial z$, which is indicative of more efficient turbulent diffusion in the vertical direction. Also, the height of the concentration profile is increased for higher $\varphi$ values. The temperature profiles (Figure \ref{fig:profilemean}g-i) show that, for a fixed distance from the source, increasing the heating intensity results in both a higher wall temperature and a thicker thermal boundary layer. 
Concentration and temperature data are combined to evaluate the vertical variation of $\overline{b}$ (Figure \ref{fig:profilemean}j-l). The latter can also be obtained by temporally averaging Equation \ref{eq:CH1bapprox}. These profiles reveal that, even close to the source,  the wall heating induces the formation of a near-wall region characterized by unstable stratification, i.e., with $\partial \overline{b} / \partial z < 0$. 
As a result, a convective boundary layer develops near the heated wall. Convective boundary layers are known to exhibit higher wall drag compared to neutrally or stably stratified cases \citep{Marucci2020}. The more unstable the boundary layer, the higher the wall drag. This explains why the near-wall phenomenology of the velocity profiles of gravity currents advancing over increasingly heated surfaces closely resembles that of gravity currents propagating over beds with increasing roughness, presenting a greater momentum loss at the wall with respect to the smooth case \citep{Sequeiros2010,Varjavand2015}, i.e., they show a reduction of the maximum velocity and a larger vertical extent of the intermediate, almost-constant velocity layer. 
Above the unstable zone, the buoyancy $\overline{b}$ attains a maximum, beyond which the flow reverts to a stably stratified configuration. In the most heated case, at $x/h_s=43$ (Figure \ref{fig:profilemean}l), negative values of $\overline{b}$ are observed near the wall, corresponding to fluid densities $\overline{\rho}$ lower than the ambient reference density $\rho_0$. 
\begin{figure}[!h]
\centering
\includegraphics[width=1\textwidth]{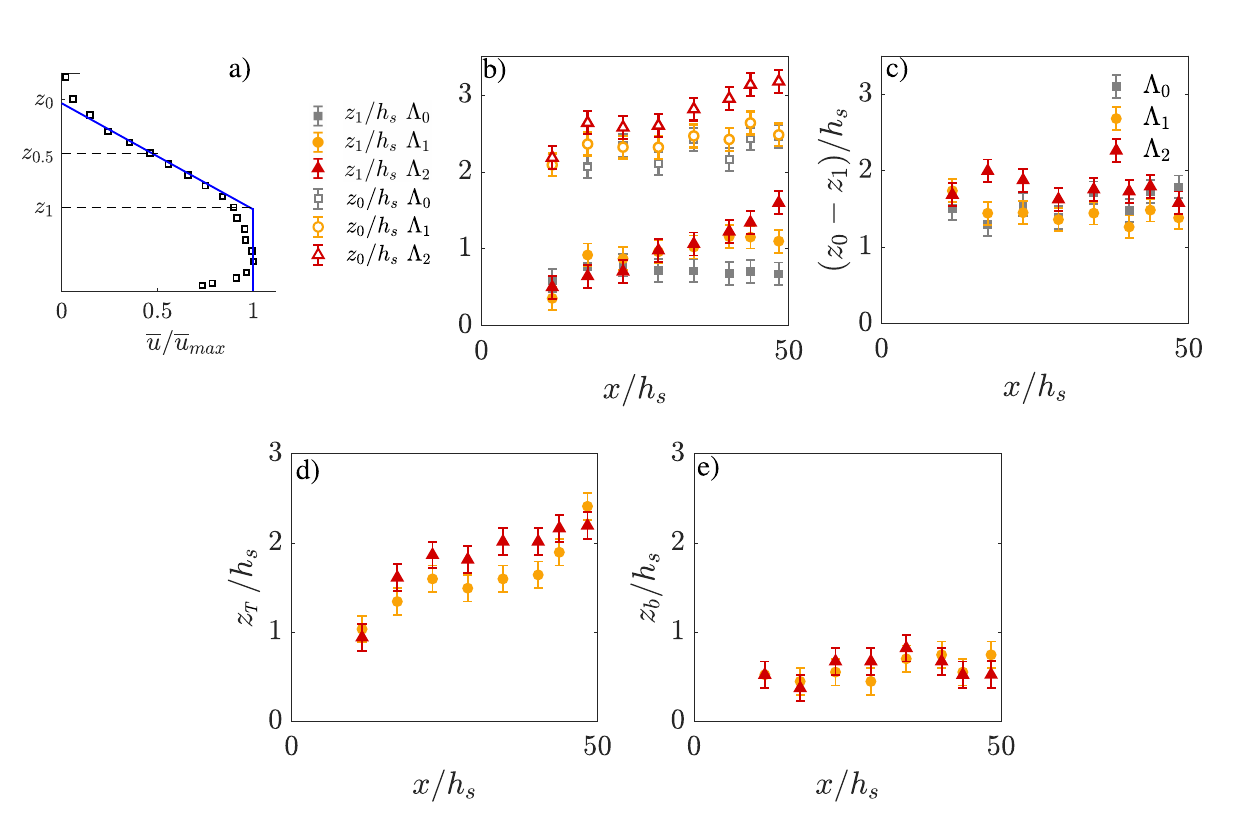}
\caption{Longitudinal variation of different characteristic heights of the currents. Panels (a) and (b) show the definition and the values of the height characterising the velocity profile,$z_1$ and $z_0$,  defined as \cite{Odier2014}. Panel (c) shows the thickness of the mixing zone. Panels (d) and (e) show the thickness of the thermal boundary layer, $z_T$, and the position of the maximum of buoyancy, $z_b$, respectively.}
\label{fig:heights}
\end{figure}

To provide a comprehensive view of the longitudinal variation of the current under different heating conditions, different characteristic current heights are plotted as a function of the longitudinal coordinate in Figure \ref{fig:heights}. To characterize the structural development of the velocity profile, we adopt the same definition of characteristic heights as in \cite{Odier2014} (Figure \ref{fig:heights}a). The centre-line of the mixing zone, denoted as $z_{0.5}(x)$, is defined by the condition $\overline{u}(x, z_{0.5}) = 0.5\,\overline{u}_{max}$, where $\overline{u}_{max}(x)$ is the maximum streamwise velocity. The lower ($z_1$) and upper ($z_0$) boundaries of the mixing zone are determined by the intercepts of the linear curve fitting of $\overline{u}$ at $z_{0.5}$ with the maximum velocity, $\overline{u}=\overline{u}_{max}$, and with the vertical axis, $\overline{u}=0$, respectively. The longitudinal variation of $z_1$ and $z_0$ for the three heating conditions is shown in Figure \ref{fig:heights}b. The value $z_{0.5}$ is not shown, but it can be easily obtained as the midpoint between $z_1$ and $z_0$. The error bars correspond to $\pm 1$~cm, which matches the vertical spacing of the LDV measurements within the mixing region. Differences between the various heating cases are minimal for $x < 20h_s$, but become more pronounced in the second half of the channel. 
The values of $z_1$ indicate that the height of the almost-constant velocity zone increases with increasing heating, i.e., with increasing wall drag. Consequently, both $z_{0.5}$ and $z_0$ also increase.
An interesting result emerges when plotting the thickness of the mixing zone, defined as $z_0 - z_1$ (Figure \ref{fig:heights}c): no clear trend is observed in the thickness of this region as heating increases. This indicates that the increase due to heating in the total height of the velocity profile, $z_0$, is primarily attributed to the expansion of the almost-uniform velocity region, rather than to an increase in the thickness of the mixing zone.  
Figures \ref{fig:heights}d,e show the thickness of the thermal boundary layer $z_{T}$ and the height $z_b$ of the point of inversion of the stability condition, i.e., where $\partial \overline{b}/\partial z=0$. The thickness of the boundary layer is defined as the height where $\overline{\Delta T}<0.1 \overline{\Delta T} _{bmax}$, where $\Delta T_{bmax}$ is the temperature difference where the buoyancy is maximum.  This definition of boundary layer thickness differs from the classical one, in which the reference $\overline{\Delta T}$ is taken at the wall \citep{cengel_2010}. However, with our measurement system, it is challenging to obtain an accurate estimate of the wall temperature. 
Since the temperature gradient reaches its maximum at the wall, a vertical positioning uncertainty of $1$~mm for the probe can introduce significant errors in the estimate of the wall temperature. Furthermore, due to the fragility of the cold wire, the lowest accessible temperature measurement point is located 3--5~mm above the wall.
At all $x$-locations where measurements were performed, the thickness of the thermal boundary layer exceeds the extent of the almost-uniform velocity region, $z_1$ (Figures \ref{fig:heights}d). The position of the buoyancy maximum (Figures \ref{fig:heights}e), $z_b$, remains confined below $z_1$, showing that, for all the investigated conditions, the unstably stratified region is maintained close to the wall and does not extend into the mixing region.  \\
\begin{figure}[!h]
\includegraphics[width=1\textwidth]{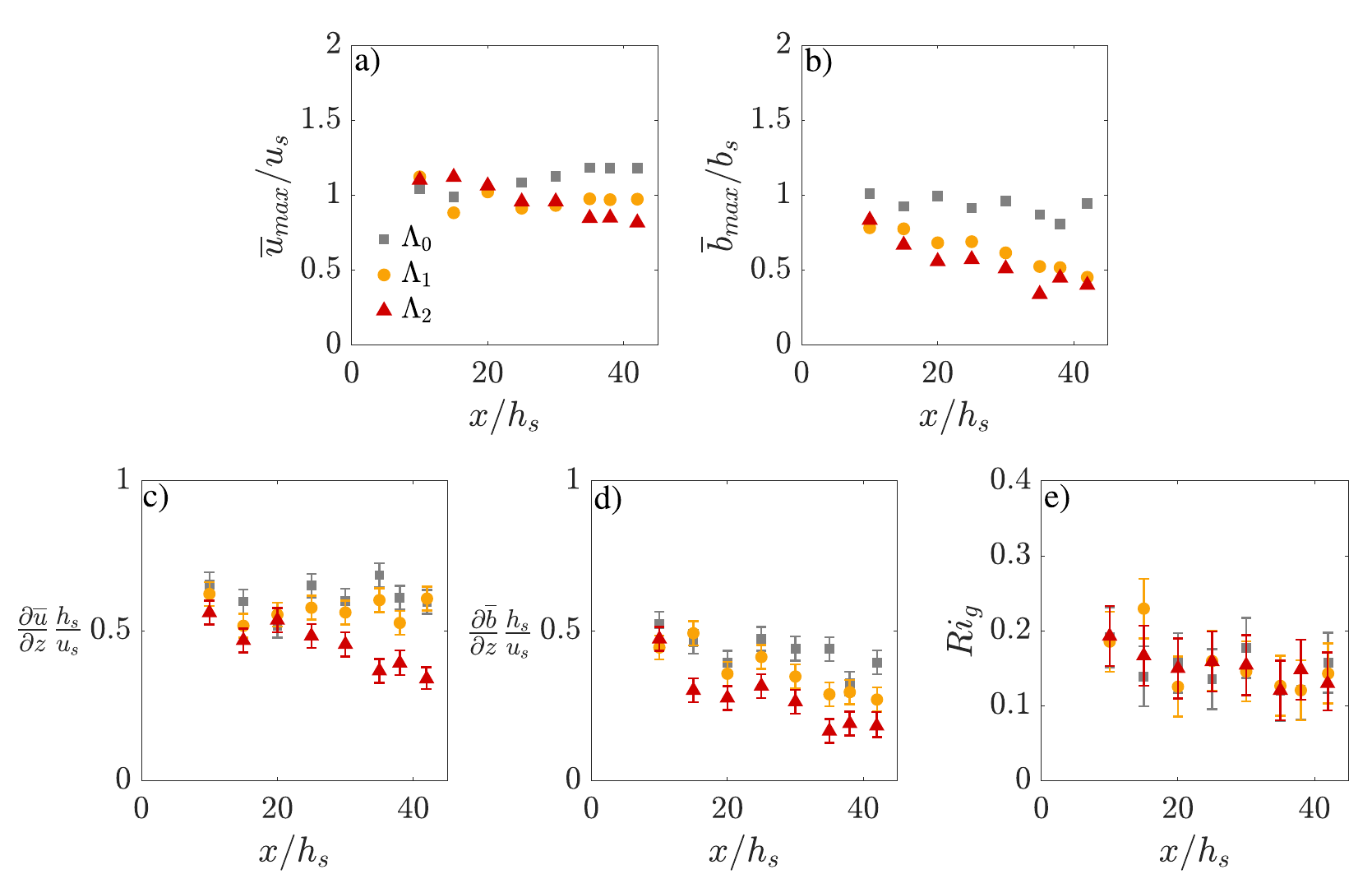}
\caption{Longitudinal evolution of mean horizontal velocity maximum (a) and mean buoyancy maximum (b). Panels (c) and (d) present the non-dimensional shear and vertical derivative of buoyancy averaged over the mixing zone. Panel (e) presents the longitudinal variation of the gradient Richardson number, averaged over the mixing zone.}
\label{fig:Rig}
\end{figure}
Figure \ref{fig:Rig}a,b shows the longitudinal evolution of the horizontal velocity and buoyancy maxima, $\overline{u}_{max}$ and $\overline{b}_{max}$, respectively. For $\Lambda_0$, the velocity becomes almost constant in the last part of the channel, showing that it is approaching the normal condition \citep{Martin_Negretti_Hopfinger_2019}, while it still decreases for $\Lambda_1$ and $\Lambda_2$ (Figure \ref{fig:Rig}a). The reduction of $\overline b_{max}$ is almost linear with $x$ for both $\Lambda_1$ and $\Lambda_2$ (Figure \ref{fig:Rig}b). For $\Lambda_1$ and $\Lambda_2$, the longitudinal development of the mixing region is characterized by the interplay of the vertical shear, $S=\partial \overline{u}/\partial z $ and the stratification $N^2=\partial \overline{b}/ \partial z $, being $N$ the Brunt-Vaisala frequency. The mixing with the external air is produced by an instability whose strength depends on the gradient Richardson number, $Ri_g=N^2/S^2$ \citep{Peltier2003, Kneller2016}. Figure \ref{fig:Rig}c,d,e shows the streamwise evolution of $S$, $N^2$, and $Ri_g$, respectively, in the mixing region. All points are obtained by averaging the vertical gradients of velocity and buoyancy between $z_1$ and $z_0$, that is, along the mixing zone. The same averaging strategy was adopted in \cite{Odier2014}. As heating increases, both $\overline{u}_{max}$ and $\overline{b}_{max}$ decrease for larger $x$, while the height of the mixing region remains approximately constant. This explains the reduction in both the vertically averaged shear and stratification with increasing distance from the source in the heated case (Figure \ref{fig:Rig}c,d). Interestingly, however, the gradient Richardson number does not exhibit a clear trend across the different heating conditions and remains in the range $0.1 < Ri_g < 0.2$ for any distance from the source. These values, lower than the stability threshold of $Ri_g = 0.25$ \citep{Miles1961,Howard1961}, allow for the development of Kelvin–Helmholtz instabilities, in agreement with the observations from flow visualizations (Figure \ref{fig:CH1visual}).

\subsection{Second-order statistics}
\label{subsec:ch1secondorder}
\begin{figure}[!h]
\centering
\includegraphics[width=1.0\textwidth]{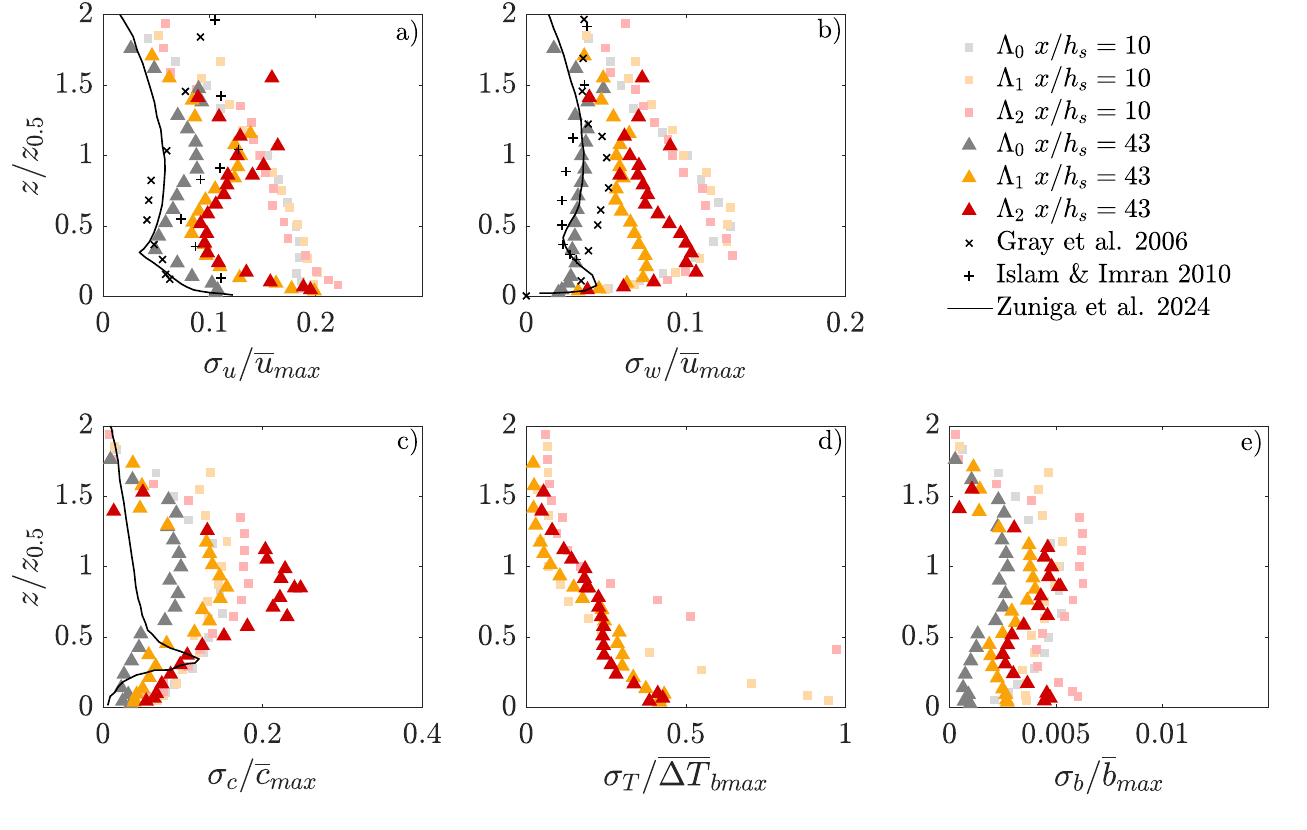}
\caption{Non-dimensional vertical profiles of standard deviation of horizontal (a) and vertical (b) velocity, CO$_2$ concentration (c), temperature (d), and buoyancy (e). 
The profiles are shown for the three heating levels at two distances from the source, $x/h_s=10$ (squares) and $x/h_s=43$ (triangles). For reference for the adiabatic case,  the $\times$ and $+$ symbols show the experimental results obtained by \cite{Gray2006} and \cite{Islam2010}, respectively, while the solid lines show the DNS results of \cite{Zuniga2024}. 
 }
\label{fig:profilesigma}
\end{figure}

The simultaneous, high-frequency measurements adopted in this study enable the evaluation of both the auto- and cross-correlations of the different measured quantities. The analysis of these statistics can provide valuable insights into the development, structure, and dynamics of heated gravity currents.  Figure~\ref{fig:profilesigma}a-e shows the standard deviation vertical profiles of the longitudinal velocity, $\sigma_u$, vertical velocity, $\sigma_w$, concentration, $\sigma_c$, temperature, $\sigma_T$, and buoyancy $\sigma_b$. The five quantities are scaled with $\overline{u}_{max}$, $\overline{u}_{max}$, $\overline{c}_{max}$ (the concentration maximum),  $\overline{\Delta T}_{bmax}$, and  $\overline{b}_{max}$, respectively. The distance from the wall $z$ is scaled with $z_{0.5}$.
The profiles for the three heating intensities are shown at two different distances from the source, $x=10h_s$ and $x=43h_s$. 
For reference for the adiabatic case, velocity statistics from the experimental studies of \cite{Gray2006} and \cite{Islam2010}, conducted on saline gravity currents propagating along adiabatic slopes of $3^\circ$ and $4.6^\circ$, respectively, are shown. Additionally, the solid line represents the DNS results from \cite{Zuniga2024} in the near self-similar regime of a current traveling on an adiabatic, 3$^\circ$ slope. At $x=10h_s$, the $\sigma_u$, $\sigma_w$, and $\sigma_c$ values are comparable across all three heating conditions. However, differences become evident further downstream (Figure \ref{fig:profilesigma}a-c).
In particular, in the unheated case, since the slope in the experiment is only of $3^\circ$, stratification effectively damps turbulent fluctuations, leading to a reduction in turbulence intensity with distance from the source. This damping effect is less pronounced when energy is introduced into the system through wall heating. The decrease in fluctuation intensity of $u$, $w$, and $c$ from $x=10h_s$ to $x=43h_s$ becomes progressively smaller with increasing heating intensity.  
For all three heating configurations, the $\sigma_u$ profiles exhibit two distinct local maxima: one near the wall and another around $z\approx z_{0.5}$, while a minimum is observed at the location of maximum streamwise velocity \citep{Salinas2021,Zuniga2024} (Figure \ref{fig:profilesigma}a). The r.m.s of the vertical velocity fluctuations, $\sigma_w$, is smaller than that of the horizontal component, $\sigma_u$, throughout the vertical profile (Figure \ref{fig:profilesigma}b).
The profile of $\sigma_c$ has a maximum at $z\approx z_{0.5}$, while $\sigma_T$ is maximum at the wall (Figure \ref{fig:profilesigma}d).   
The $\sigma_b$ profiles (Figure \ref{fig:profilesigma}d) depend on the concentration and temperature fluctuations \ref{eq:CH1bapprox}.
In the non-heated case, the $\sigma_b$ and $\sigma_c$ profiles are equivalent. In contrast, for the heated cases, the $\sigma_b$ profiles exhibit two distinct local maxima: the first occurs near the wall and is associated with temperature fluctuations, while the second is around $z \approx 0.9\, z_{0.5}$ and is mainly associated with fluctuations in concentration.

\begin{figure}[!h]
\centering
\includegraphics[width=0.89\textwidth]{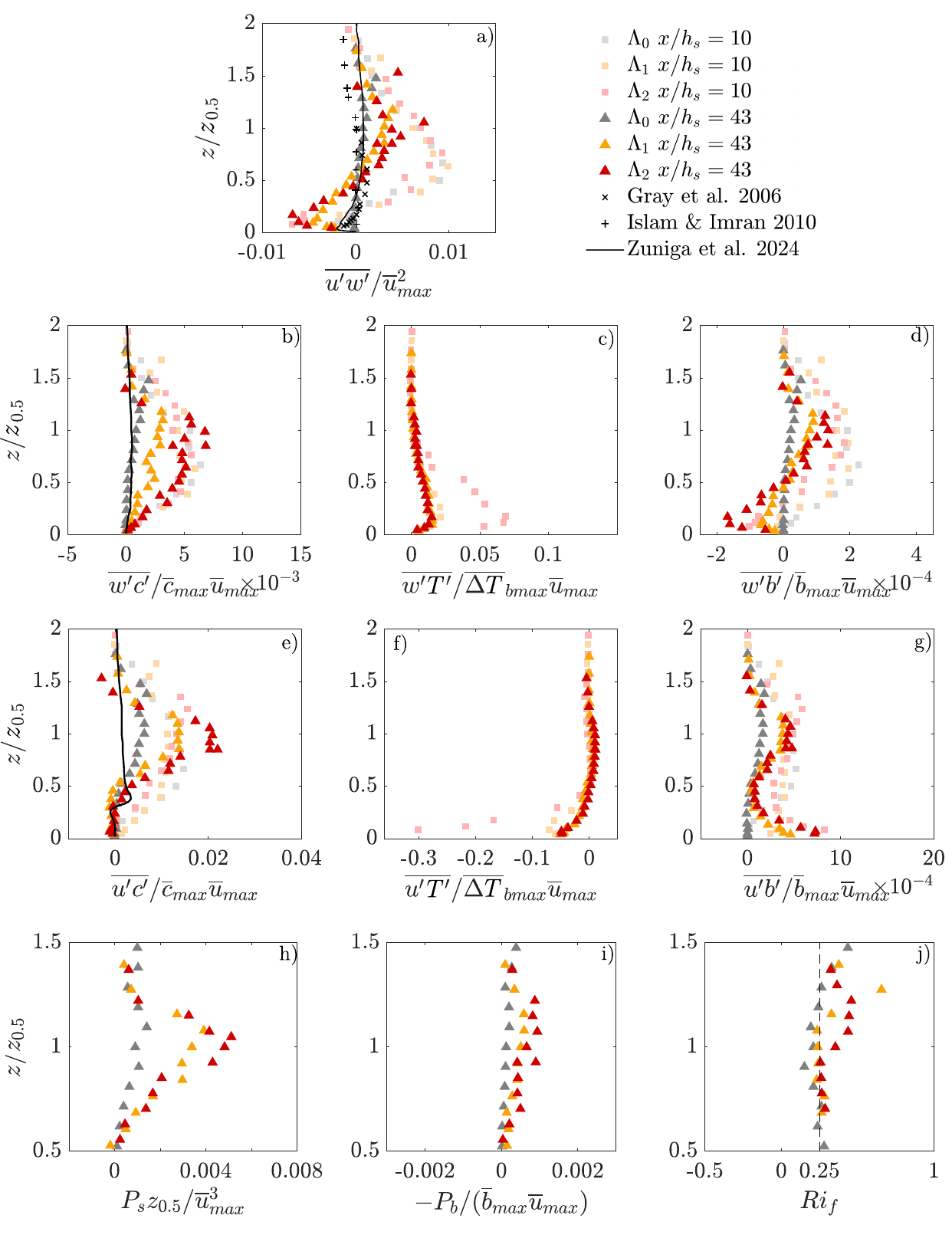}
\caption{Panels (a-g) show the cross-correlation of the measured quantities for the three heating levels considered (a)-(g). Panels (h) and (i) show the non-dimensional production of turbulent kinetic energy due to shear and buoyancy, respectively. The flux Richardson number is shown in panel (j). }
\label{fig:CH1profilesturb}
\end{figure}
The cross-correlations between the measured quantities are shown in Figure \ref{fig:CH1profilesturb}(a-g). Again, the profiles for all the tested heating levels are shown at $x=10h_s$ and $x=43h_s$.
Figure \ref{fig:CH1profilesturb}(a) shows the Reynolds stress $\overline{u'w'}$ vertical profiles normalized with $\overline{u}_{max}^2$. At $x=43h_s$, the profiles show a negative minimum near the wall and a maximum in the mixing region. The absolute value of both peaks increases with increasing heating intensity. The growth in the negative peak near the wall is attributed to the presence of the unstable boundary layer, which is characterized by larger absolute values of $\overline{u'w'}$ for higher levels of instability (i.e., larger $\partial b/\partial z$) \citep{Marucci2020}. The near-wall $\overline{u'w'}$ values reinforce the discussion presented in Section \ref{subsec:CH1firstorder} regarding the enhancement in wall drag with increasing heating. Indeed, the drag coefficient $C_d$ can be related to the near-wall Reynolds stresses by the relation $-\overline{\rho}\overline{u'w'}=1/2 \overline{\rho} C_d \overline{u}^2_{max} $. With this definition, the near-wall values plotted in Figure\ref{fig:CH1profilesturb}a, whose norm grows with $\Lambda_s$, are equal to $-C_d/2$.
The growth in the mixing region of $\overline{u'w'}$ with heating reflects the same trend observed in $\sigma_u$ and $\sigma_w$, and is attributed to the reduced stratification-induced damping of turbulence in the heated configurations compared to the adiabatic case.

The wall-normal and wall-tangent turbulent fluxes of concentration, temperature, and buoyancy are shown in Figure \ref{fig:CH1profilesturb}(b-g). 
From Equation \ref{eq:CH1bapprox}, the turbulent buoyancy fluxes can be written as a function of the concentration and temperature ones as:
\begin{align}
    \overline {w'b'} &\approx  g \biggl(\chi_M \overline {w'c'} - \frac{\overline {w'T'}}{T_0}\biggr) & \text{and} & &  \overline {u'b'} &\approx  g \biggl(\chi_M \overline {u'c'} - \frac{\overline {u'T'}}{T_0}\biggr).\label{eq:CH1turbufluxes}
\end{align}
The values of $\overline{w'c'}$ and $\overline{w'T'}$ are positive throughout the entire channel height. In contrast, the buoyancy flux ($\overline{w'b'}$, Figure \ref{fig:CH1profilesturb}d) changes sign in the heated experiments (Figure \ref{fig:CH1profilesturb}d). For $z < 0.5\, z_{0.5}$, the turbulent buoyancy flux is negative, indicating an upward turbulent transport of light fluid (with negative buoyancy) induced by the unstable stratification. Notably, a positive $\overline{w'T'}$ corresponds to a negative contribution to $\overline{w'b'}$ (Equation \ref{eq:CH1turbufluxes}). For $z > 0.5\, z_{0.5}$, the value of $\overline{w'b'}$ becomes positive.
The turbulent horizontal fluxes of CO$_2$ concentration, $\overline{u'c'}$, are positive for all the considered heating conditions (Figure \ref{fig:CH1profilesturb}e).
This marks a difference compared to what is observed in neutral and dense plumes released in a turbulent boundary layer, where the value of $\overline{u'c'}$ is negative throughout the entire plume \citep{Marro2020,Vidali2022}. The results obtained in this work are consistent with those of \cite{Zuniga2024}, who showed that for gravity currents propagating over an adiabatic inclined wall, the value of $\overline{u'c'}$ is positive and increases for steeper slopes.
The $\overline{u'T'}$ term (Figure \ref{fig:CH1profilesturb}f) is negative near the wall and contributes to the positive peak observed in this region of the horizontal turbulent buoyancy flux $\overline{u'b'}$ (Figure \ref{fig:CH1profilesturb}g).

The second-order statistics profiles can be exploited to compute the flux Richardson number, $Ri_f$.
The latter can be defined as $Ri_f=-P_b/P_s$ \citep{vanReeuwijk2019}, where:
\begin{align}
P_b&=\overline{u'b'}sin (\phi) -\overline{w'b'}cos(\phi)\approx -\overline{w'b'} & &\mathrm{and} & P_s&=-\overline{u'w'}\frac{\partial\overline u}{\partial z}    
\end{align}
are the buoyancy and shear production of turbulent kinetic energy, respectively. 
Although $Ri_f$ does not rigorously represent mixing, it is often referred to as the mixing efficiency \citep{Ivey1991,Caufield_2000,Pardyjak2002}. The non-dimensional values of $P_s$ and $-P_b$ are shown in Figure~\ref{fig:CH1profilesturb}h,i. To focus on the mixing region,  the results are shown only for $0.5<z/z_{0.5}<1.5$. Heating leads to an increase in both the norm of $P_s$ and $P_b$. Interestingly, however, the flux Richardson number does not exhibit a clear trend with heating and remains approximately $0.25$ (Figure~\ref{fig:CH1profilesturb}j).

\subsection{Integral balances and measurements validation}
\label{subsec:CH1integrals}

To assess the reliability of our measurements, the wall-normal profiles shown in Sections \ref{subsec:CH1firstorder} and \ref{subsec:ch1secondorder} can be integrated to evaluate the balances of CO$_2$ mass, enthalpy, and buoyancy at different distances from the source.  
We start presenting the analytical formulation of these fluxes,  assuming that the flow is homogeneous in the spanwise ($y$) direction.

The integral longitudinal (i.e., parallel to the wall)  flux of CO$_2$ mass is defined as 
$G=\int_0^{\infty} \overline{\rho u \tilde{c}} \, \D z$ and must be constant and equal to the value at the source $G_s=\rho_s h_s u_s \tilde c_s$ for all the streamwise positions $x$ considered. 
Exploting the Boussinesq approximation (i.e., imposing $\rho=\rho_0$), and Equation \ref{eq:massfraction} to link the mass fraction $\tilde{c}$ with the volume fraction $c$, $G/G_s$ can be written as:
\begin{align}
    \frac{G}{G_s}\approx \frac{\rho_0}{\rho_sh_s u_s c_s }\int_0^{\infty} \overline{u c} \, \D z = \frac{\rho_0}{\rho_sh_s u_s c_s } \bigg (\int_0^{\infty} \overline{u}\ \overline{c} \, \D z +\int_0^{\infty} \overline{u'c'} \, \D z \bigg)=  \frac{\overline{G}+G'}{G_s}=1.\label{eq:CH1massbalance}
\end{align}
The total CO$_2$ mass flux $G$, is given by the sum of the mean flow contribution, $\overline{G}$, obtained with the integral of $\overline{u} \,\overline{c}$, and the contribution of the turbulent flux $G'$.

The balance equation for the enthalpy excess with respect to the enthalpy of ambient air ($h_0 = c_{p,0} T_0$) can be expressed as
\begin{equation}
\frac{\D H}{\D x}  = \varphi, \label{eq:cons:FH}
\end{equation}
with $H = \int_0^\infty \overline{\rho ( c_p T - c_{p,0} T_0)u} \, \D z$. Similarly to what was done for $G$, this definition can be simplified by applying the Boussinesq approximation and assuming $c_{p} = c_{p,0}$ (Equation \ref{eq:cp}). The contributions of the mean field and of the turbulent fluctuations can be separated as:
\begin{align}
    H\approx \rho_0 c_{p,0}\int_0^{\infty} \overline{u (\Delta T)} \, \D z = \rho_0 c_{p,0}\int_0^{\infty} \overline{u}\,\overline{\Delta T} \, \D z +\rho_0 c_{p,0}\int_0^{\infty} \overline{u'T'} \, \D z = \overline{H} +H'.\label{eq:enthalpydefinition}
\end{align}
In our experiment, the temperature of the mixture at the source is assumed equal to the ambient air temperature far from the wall, $\Delta T_s=0$,  and the value of $\varphi$ is constant in the $x$ direction. Therefore, the integration of Equation \ref{eq:cons:FH} in the streamwise direction leads to:
\begin{align}
    H=\overline{H}+H'=\varphi x.\label{eq:CH1enthalpybalance}
\end{align}
Finally, based on the relation \ref{eq:CH1bapprox}, the flux of buoyancy parallel to the wall, $F$, is written as 
\begin{equation}
F \equiv  \int_0^{\infty} \overline{ b u} \, \D z 
\approx g \int_0^{\infty} \overline{ \left ( \chi_M c - \frac{\Delta T}{T_0} \right ) u} \, \D z
= \overline{F} + F', \label{eq:def:Buoyancyflux}
\end{equation}
with $\overline{F} = g  \int_0^{\infty} 
 \overline{u} \, (\chi_m \overline{c} - \overline{\Delta T}/T_0 )\, \D z  $ 
and
$F' = g  \int_0^\infty (\chi_m \overline{u' c'} - \overline{u' T'}) \, \D z $. 
Using the definitions of $G$ and $H$, $F$ can be expressed as 
a linear combination of the CO$_2$ mass and enthalpy fluxes:
\begin{equation}
F= \frac{\chi_M g}{\rho_0}\frac{M_0}{M_{\text{CO}_2}}  G 
- \frac{g }{\rho_0 c_{p,0} T_0 }H . \label{eq:FB}
\end{equation}
Since $G$ is constant (Eq. \ref{eq:CH1massbalance}) and $H$ increases linearly with $x$ (Eq. \ref{eq:CH1enthalpybalance}), the streamwise derivative of F is:
\begin{equation}
\frac{\D F}{\D x}  = - \frac{\varphi g }{\rho_0 c_{p,0} T_0 } . \label{eq:cons:FB}
\end{equation}
Integrating Eq. \eqref{eq:cons:FB} and dividing by the buoyancy flux at the source $F_s=b_su_sh_s$, leads to the final formulation of the buoyancy flux value as a function of the distance from the source:
\begin{equation}
\frac{F}{F_s} = 1 -  \Lambda_s \frac{x}{h_s}. \label{eq:CH1buoyancybalance}
\end{equation}
This equation clarifies the meaning of the non-dimensional parameter $\Lambda_s$ (Equation~\ref{eq:ch1lambda}) used to quantify the heating intensity. It represents the constant rate of decrease of the non-dimensional buoyancy flux $F/F_s$ as a function of the distance from the inlet.

\begin{figure}
\centering
\includegraphics[width=0.93\textwidth]{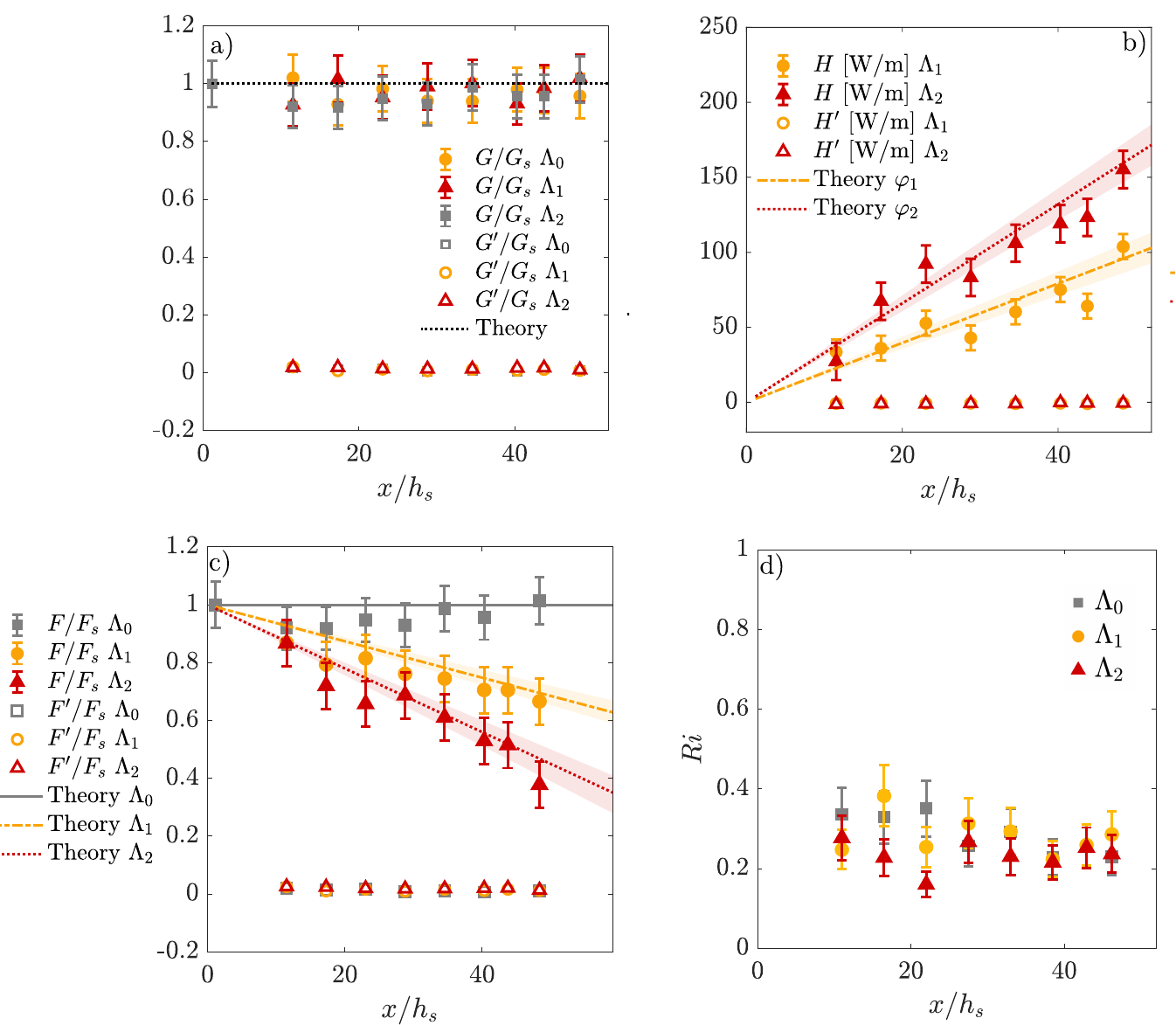}
\caption{Longitudinal evolution of the total and turbulent fluxes of CO$_2$ mass $G$(a), excess of enthalpy with respect to the source condition $H$ (b), and buoyancy $F$(c). The theoretical values of the fluxes (Eq. \ref{eq:CH1massbalance}, \ref{eq:CH1enthalpybalance}, \ref{eq:CH1buoyancybalance})  are also plotted. Panel (d) shows the bulk Richardson number obtained with Equation \ref{eq:CH1Ri}.}
\label{fig:balances}
\end{figure}
Figure \ref{fig:balances}(a-c) shows the experimental values of the total and turbulent fluxes of CO$_2$ mass, excess of enthalpy, and buoyancy, respectively, and a comparison with the reference Equations \ref{eq:CH1massbalance}, \ref{eq:CH1enthalpybalance}, \ref{eq:CH1buoyancybalance}, for the different heating considered.  The CO\textsubscript{2} mass balance (Figure \ref{fig:balances}a) shows that our results remain within a 10\% error margin relative to the constant mass flux value imposed at the source. These discrepancies are consistent with those reported in previous mass balance studies conducted by \cite{Marro2020} and \cite{Vidali2022}, with passive scalar and CO$_2$, using the same instrumentation, to evaluate pollutant emissions from a line source and a point source, respectively, in wind tunnel experiments.
The values of $G'$ are positive and their contribution to the total flux remains below $2\%$ for all the heating intensities considered. Consequently, in absolute values, the largest uncertainty in the estimate of the total flux $G$ is associated with the mean term $\overline{u}\, \overline{c}$ \citep{Raupach_Legg_1983,Marro2020,Vidali2022}. 

The experimental results for the enthalpy and buoyancy fluxes (Figure \ref{fig:balances}b,c) show good agreement with the theoretical predictions (Eqs.~\ref{eq:CH1enthalpybalance} and \ref{eq:CH1buoyancybalance}). In both cases, the contribution of turbulent fluxes to the total ones is negligible. The shaded areas enclosing the theoretical curves represent the uncertainty on the estimation of $\varphi$ and $\Lambda_s$ (Section \ref{sec:CH1inlet}) related to the possible values of the emissivity. Specifically, they are obtained by assuming an emissivity between 0.9 and 1. 
The agreement between the experimental fluxes and the theoretical relations demonstrates the reliability of the simoultaneous LDV–FID–CCA measurements adopted in this study and the accuracy of the estimated convective heat flux $\varphi$, as well as the effectiveness of our experimental setup in ensuring (i) a homogeneous heating along the channel floor and (ii) uniform flow conditions in the $y$-direction, sufficiently far from the lateral walls. Figure \ref{fig:balances}c also provides a quantitative indication of the heating intensity relative to the initial buoyancy flux at the source, $F_s=u_sb_sh_s$. In the weakly heated case $\Lambda_1$, the buoyancy flux at the end of the channel reaches 60\% of its initial value, while in the strongly heated case $\Lambda_2$, it decreases to 35\% of $F_s$.

Finally, the $z$-integrated statistical profiles can also be exploited to compute the bulk Richardson number, $Ri$. One possible definition of this quantity is:
\begin{align}
    Ri=\frac{ \int_0^{\infty} \overline{ b } \, \D z \bigg(\int_0^{\infty} \overline{ u} \, \D z \bigg)^2 \cos\phi}{\bigg(\int_0^{\infty} \overline{  u}^2 \, \D z \bigg)^2}. \label{eq:CH1Ri}
\end{align}
The experimental values of $Ri$ for the three heating intensities considered are shown in Figure \ref{fig:balances}d.
Notably, consistently with the values obtained for $Ri_g$ and $Ri_f$, no clear trend of $Ri$ with increasing wall heating is observed. This indicates that, along the entire channel length, the flow remains in a supercritical regime, i.e., with $Ri<1$, preventing the outlet conditions from influencing the whole flow domain and thus justifying the choice of imposing a $3^\circ$ slope.

\section{Conclusion}\label{sec:CH1conclusion}
We have presented a laboratory experiment on sustained gravity currents propagating along a slightly inclined, heated wall. The dense current is generated using a mixture of air and CO$_2$.
Despite the large number of experiments on gravity currents, to the best of our knowledge, no laboratory studies on steady gravity currents propagating along non-adiabatic boundaries, typical of numerous lower-atmosphere flows, are available in the literature. 
To obtain a complete point-wise characterization of the flow, simultaneous measurements of velocity (Laser Doppler Velocimetry), CO$_2$ concentration using C$_2$H$_6$ as tracer (Flame Ionization Detector), and temperature (cold-wire) are performed.  
The effects related to the presence in the mixture of CO$_2$ and seeding droplets on the FID and CCA measurements are evaluated.
We presented the experimental protocol adopted to reconstruct the density signal and estimate the turbulent fluxes of CO$_2$, temperature, and buoyancy.  
The reliability of our experimental procedure, as well as the estimate of the effective heat flux exchanged with the current and its spatial uniformity, is assessed through integral balances of CO$_2$ mass, excess enthalpy with respect to the source condition, and buoyancy fluxes, performed at various distances downstream of the source.  \\
The current buoyancy flux, normalized with its value at the source, $F_s$, decreases linearly with the distance from the source $x/h_s$, at a rate equal to $\Lambda_s$, the ratio between the wall-normal buoyancy flux per unit area induced by the heating $\varphi$, and the streamwise buoyancy flux per unit area that generates the current, $u_sb_s$. \\
The first- and second-order statistics vertical profiles for three different heating conditions, namely, an adiabatic case ($\Lambda_0$) and two increasingly heated configurations ($\Lambda_1$ and $\Lambda_2$) are evaluated. The main observations are summarized below. 
\begin{enumerate}

    \item Wall heating induces the development of an unstable convective boundary layer, which enhances the vertical turbulent transport near the wall of both CO$_2$ and temperature, for increasing $\Lambda_s$.
    
    \item The overall height of the velocity profiles increases with both wall heating and downstream distance from the source, while the maximum horizontal velocity decreases for increasing $\Lambda_s$. The vertical profile growth is primarily associated with the thickening of the near-wall region characterized by an approximately constant velocity. In contrast, the thickness of the mixing region, defined as the zone with a nearly constant velocity gradient, does not show a clear dependence on heating.
    
    \item Within the mixing region, both the vertical shear $S$ and the Brunt–Väisälä frequency $N$ (averaged over the height of the mixing layer) are reduced as $\Lambda_s$ increases. However, the gradient Richardson number $Ri_g = N^2/S^2$, does not show a trend with heating intensity.
    

    \item In the second half of the channel, i.e., $x>25h_s$, the standard deviation $\sigma$ of all measured quantities increases over the entire height of the gravity current for higher $\Lambda_s$. While for the adiabatic case $\Lambda_0$, the fluctuations are attenuated compared to those measured at $x/h_s = 10 h_s$, the heating can reduce or eliminate this attenuation.

    \item In the mixing region, both norms of the shear production $P_s$ and the buoyancy production $P_b$ of turbulent kinetic energy increase with $\Lambda_s$. However, the ratio $-P_b/P_s$, i.e., the flux Richardson number, does not show a clear trend with the heating intensity. 

    \item The bulk Richardson number does not show a clear trend with the imposed $\Lambda_s$.
\end{enumerate}

The results obtained in this study can serve as a valuable benchmark for the validation of numerical models such as LES (Large Eddy Simulation), RANS (Reynolds-Averaged Navier–Stokes), or simplified models aimed at simulating atmospheric gravity currents along heating surfaces. Validated numerical simulations can then be employed to extend the $\Lambda_s$ and slope space beyond the experimental conditions. 
In particular, simulations can also be employed to investigate the development of the currents over distances greater than those analyzed here, ideally up to the point where the buoyancy flux F becomes zero. It will be interesting to assess whether the findings related to the Richardson numbers, i.e., a substantial independence of $Ri_g$, $Ri_f$, and $Ri$ on the heating condition, presented in this work, remain valid up to the complete vanishing of the longitudinal buoyancy flux.

\singlespacing

\section*{Author contributions}
All authors contributed to the design of the facility, and A.Z. realized the final project. S.L. performed the experiments, analyzed the data, and wrote the manuscript. M.M. developed the experimental software, supervised the experimental campaign, and revised the manuscript. M.C. contributed to the formal analysis, supervised the research, and revised the manuscript. P.S. conceived the study, secured funding, supervised the research, contributed to the formal analysis, and revised the manuscript.\\

\section*{Acknowledgments}
The authors would like to thank the Service Étude Fabrication of LMFA for the construction of the experimental facility, and Horacio Correia for providing technical solutions to many laboratory problems. They also thank Emmanuel Jondeau for supplying the cold wires and for his assistance with LDV measurements, Loïc Mees and  Nathalie Grosjean for their valuable experimental advices, and Samuel Vaux for fruitful discussions.

\section*{Conflict of Interest Statement}
The authors report no conflict of interest.

\section*{Funding}
This work was founded by the Région Auvergne-Rhône-Alpes (France) within the COVENTU Project (grant number: GFC
2020–043) 

\section*{Data Availability Statement}
The data presented in this study are available on request from the corresponding author.


\clearpage 
\bibliographystyle{apalike} 
\typeout{}

\bibliography{References}
\end{document}